\documentclass[review,11pt]{elsarticle}

\usepackage{lineno,hyperref}
\usepackage{amsmath}
\usepackage{color}

\usepackage{graphicx}
\graphicspath{ {images/} }
\usepackage{bm}
\usepackage[export]{adjustbox}

\makeatletter
\def\ps@pprintTitle{%
 \let\@oddhead\@empty
 \let\@evenhead\@empty
 \def\@oddfoot{}%
 \let\@evenfoot\@oddfoot}
\makeatother

\begin{document}

\begin{frontmatter}

\title{Spin Dynamics of CPMG sequence in time-dependent magnetic fields}

\author[SDR]{Martin D. H{\" u}rlimann\corref{mycorrespondingauthor}}
\ead{mhurlimann@gmail.com}

\author[SDR]{Shin Utsuzawa}
\ead{SUtsuzawa@slb.com}

\author[SDR]{Chang-Yu Hou}
\ead{CHou2@slb.com}

\cortext[mycorrespondingauthor]{Corresponding Author}
\address[SDR]{Schlumberger-Doll Research, Cambridge, MA 02139, USA}

\begin{abstract}
We analyze the effects of time dependent magnetic and RF fields on the spin dynamics of the Carr-Purcell-Meiboom-Gill (CPMG) sequence. The analysis is based on the decomposition of the magnetization into the eigenmodes of the propagator of a single refocusing cycle. For sufficiently slow changes in the external fields, the magnetization follows the changing eigenmodes adiabatically. This results in echo amplitudes that show regular modulations with time. Faster field changes can induce transitions between the eigenmodes. Such non-adiabatic behavior occurs preferentially at particular offsets of the Larmor frequency from the RF frequency where the eigenmodes become nearly degenerate. We introduce the instantaneous adiabaticity parameter ${\cal A}(t)$ that accurately predicts the crossover from the adiabatic to the non-adiabatic regime and allows the classification of field fluctuations. ${\cal A}(t)$ is determined solely by the properties of a single refocusing cycle under static conditions and the instantaneous value of the field offset and its temporal derivative. The analytical results are compared with numerical simulations.
\end{abstract}


\end{frontmatter}

\section{Introduction}

In the common implementation of NMR measurements, a significant effort is spent on ensuring that the applied magnetic field $B_0$ across the sample is both spatially and temporally as uniform as possible. However, in practice some degree of field inhomogeneities are always present. In fact, there are a number of applications where field non-uniformities are large and unavoidable. A prime example is  ex-situ NMR \cite{singlesidedNMR} that includes NMR well logging \cite{hurlimann_nmr_2015} where the NMR sensor is moved across geological samples that are located outside the  instrument.

Here we study the impact of temporal fluctuations of $B_0$ on the spin dynamics of the Carr - Purcell - Meiboom -Gill (CPMG) sequence \cite{carr54,meiboom58}. This sequence is a basic building block in many NMR measurement schemes where significant field inhomogeneities are present. Applications include sensitivity enhancement \cite{lim2002} and the monitoring of dynamic processes such as relaxation and diffusion \cite{Hurlimann2002}. The CPMG sequence is also widely used to suppress general environmental decoherence in quantum computations, with the goal towards fault-tolerance \cite{Cappellaro2006,biercuk2009}.

For complex systems with multi-exponential decays, it is generally necessary to acquire large numbers of echoes with short echo spacings, $t_E$, to cover the entire range of relaxation times. Even in simple system, it can be beneficial to acquire many echoes with short $t_E$ values, as it enables an increase in the effective signal-to-noise ratio (SNR) by averaging the response over adjacent echoes. We are here interested in the case when the field fluctuations are slow on the time scale of the echo spacing, but when the overall amplitude over the entire CPMG train can approach or exceed $B_1$, the strength of the RF field. The field fluctuations are considered to be slow when $\frac{d}{dt}B_0(t) \ll {B_1}/{t_E}$.

The carrier frequency of the rf pulses, $\omega_{rf}$, is set to fulfill the Larmor condition $\omega_{rf} = \gamma B_0$ at the beginning of the sequence. Here $\gamma$ is the gyromagnetic ratio. When $B_0$ fluctuates during the CPMG sequence, the pulses will not fulfill the Larmor condition exactly anymore. As a consequence, the refocusing pulses will not act anymore as perfect $180^\circ$ pulses in the transverse plane. Naively, one might expect that these imperfections gradually accumulate and induce a steady reduction of the echo amplitudes. However, as is clearly apparent in numerical simulations of the spin dynamics in section \ref{sec:Results}, the observed behavior is more interesting. The resulting evolution of the magnetization of the spin echoes typically shows systematic structures with  oscillations and abrupt changes. An example has been published earlier by Speier {\it et al.} \cite{speier99}.

The purpose of this paper is to develop the underlying theory and explain this behavior. The theory is presented in section \ref{sec:Theory}. This is followed by section  \ref{sec:Results} where we present numerical simulations that are compared with the theoretical results.

The theoretical analysis in section  \ref{sec:Theory} proceeds in two steps. We first derive the average Hamiltonian that describes a single refocusing cycle for an arbitrary off-resonance condition and obtain the eigenmodes. In the second step, we consider the response under this average Hamiltonian when it becomes time-dependent.  In the adiabatic regime applicable to sufficiently slow fluctuations, the evolution of the magnetization follows the eigenmodes of the instantaneous average Hamiltonian. Simple analytical expressions are obtained for this regime. We then derive the condition for the validity of the adiabatic regime. We introduce the instantaneous adiabaticity parameter ${\cal A}(t)$ that quantitatively predicts the extent of the adiabatic regime and the occurrence of non-adiabatic events. It is found that non-adiabatic events are typically confined to small regions of offset frequencies where the field fluctuations are able to induce transitions between the eigenmodes. In section \ref{sec:B1motion}, the analysis is generalized to the treatment of fluctuations in the amplitudes of both $B_0$ and $B_1$.

In section \ref{sec:Results}, we compare the analytical results obtained in section \ref{sec:Theory} with numerical simulations of the CPMG response in non-stationary $B_0$ fields. We demonstrate that the theoretical treatment is rather general and applicable to various forms of field fluctuations. We present detailed comparison between theory and numerical simulation for linear ramp, harmonic motion, and bi-linear field variations. We also study the dependence on the amplitude and rate of change of the fluctuations.

\section{Theory}
\label{sec:Theory}
\subsection{Introduction}
\begin{figure}
	   \centering
        \adjincludegraphics[width=5.5in, trim={{.10
       \width} {.15\width} {.00\width} {.05\width}},clip]
       {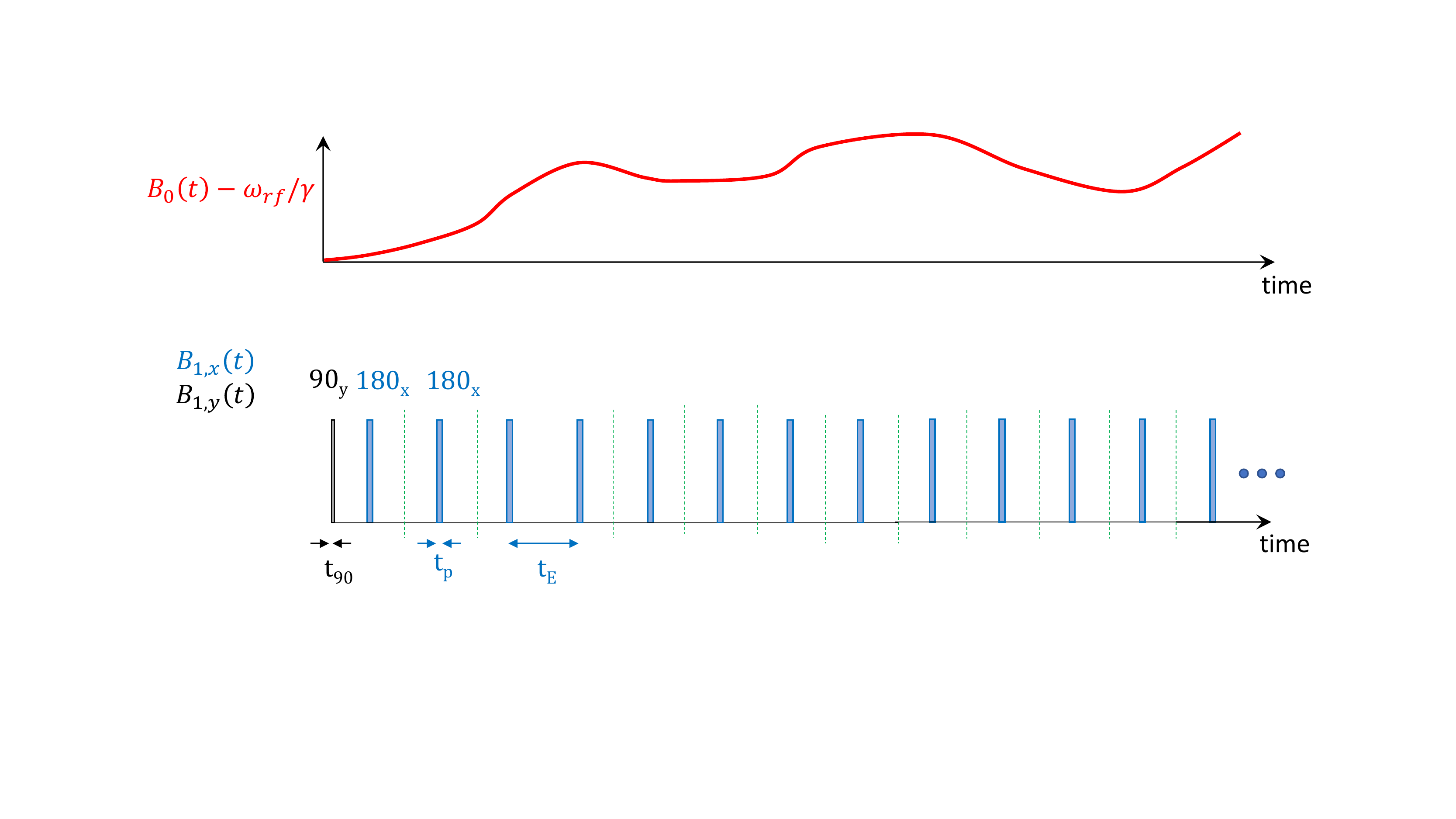}
  	 \caption[Figure 1]{Basic CPMG sequence with time-dependent $B_0$ field. The refocusing pulses are shown in blue and are assumed to have phase $0$ and duration $t_p$.  The excitation pulse with phase $\pi/2$ is shown in black and has a duration of $t_{90}$. The dotted lines show the positions of the nominal echo centers, separated by the echo spacing $t_E$.}
	   \label{fig:CPMG_varB0}
\end{figure}

In this section, we develop the theory that describes the response of the CPMG sequence under a time varying magnetic field, $B_0(t)$, as shown in Fig.\,\ref{fig:CPMG_varB0}. After the initial excitation pulse, the pulse sequence consists of a long string of rf pulses of duration $t_p$ that are equally spaced by the echo spacing $t_E$. We are interested in the resulting magnetization ${\vec M}_k$ at the nominal echo locations $t_k = k t_E$. We make the assumption that we deal with non-interacting spin 1/2 nuclei and for simplicity, we neglect $T_1$ and $T_2$ relaxation.

The relevant Hamiltonian in the rotating frame of the carrier frequency of the rf pulses, $\omega_{rf}$, is given by:
\begin{equation}
\mathcal{H} = -\hbar \gamma \left[ \left(  B_0(t) - \omega_{rf}/\gamma \right) I_z + B_{1,x}(t) I_x + B_{1,y}(t)I_y \right].
\label{eq:Hamiltonian_orig}
\end{equation}
Both the longitudinal and transverse terms are time dependent, but on a different time scale. As mentioned in the introduction, we assume that the field $B_0$ varies only slowly compared to the echo spacing, $t_E$. In contrast, the rf terms represent short pulses of duration $t_p$ that are repeated many times with a period of $t_E$. This separation of time scales allows us to proceed in two steps. We first deal with the fast time dependence of the rf pulses using average Hamiltonian theory \cite{haberlenwaugh1968,ernst} and then consider the impact of the slower variation of $B_0(t)$ on this average Hamiltonian.

In the first step, presented in section \ref{sec:averageHamiltonian}, we ignore the time dependence of $B_0$ and consider only the effect of the periodically applied refocusing pulses. The average Hamiltonian theory is well suited for this analysis. The Hamiltonian is periodic in time and also fulfills the requirement that  the generated signal is stroboscopically sampled between the refocusing pulse, synchronized with the time dependence of the Hamiltonian.  The average Hamiltonian theory  provides an analysis of the echo-to-echo propagator and has been used previously for the study of the CPMG sequence with large static field inhomogeneities \cite{hurlimann_CPMG}. We give analytical solutions for the average Hamiltonian for an arbitrary value of $B_0$ and diagonalize it. This allows us to expand the magnetization into three eigenstates.

The time dependence of $B_0(t)$ is then considered in section \ref{sec:LandauZener}. This can be viewed as a generalized treatment of the Landau-Zener problem \cite{Landau1932,Zener1932}. The dynamics of $B_0(t)$ has a minor impact on the instantaneous eigenstates of the average Hamiltonian derived for the static $B_0$ and the corrections are derived to first order. We then show that for sufficiently slow variations of $B_0(t)$, the magnetization follows the instantaneous eigenstates adiabatically and we obtain analytical solutions for this case. Next, we derive the condition for adiabaticity and introduce the adiabaticity parameter $\cal A$. We show that this parameter has a strong sensitivity on the instantaneous offset frequency of the Larmor frequency from the rf frequency, $ \gamma B_0(t) - \omega_{rf}$. Non-adiabatic events that cause transitions between the eigenstates are typically confined to small regions of offset frequency.

\subsection{Average Hamiltonian}
\label{sec:averageHamiltonian}

In the first step of the analysis, we derive the average Hamiltonian $\mathcal{H}_{\mathrm ave}$ for arbitrary (but time independent) values of $B_0$ and $B_1$.

It is useful to start the analysis by listing the key parameters that control the spin dynamics of this problem. There are three important frequencies:
\begin{enumerate}
                   \item $\omega_{rf}$, the carrier frequency of the rf pulses that is under control of the experimenter.
                   \item $\omega_0(t) \equiv \gamma B_0(t) - \omega_{rf}$, the deviation of the instantaneous Larmor frequency from $\omega_{rf}$ that is proportional to the applied field.
                   \item $\omega_1 \equiv \gamma B_{1,\perp}/2$, the nominal nutation frequency of the refocusing pulses that is proportional to the amplitude of the applied linearly polarized rf field, $B_{1,\perp}$.
\end{enumerate}

It is convenient  to normalize the frequencies with the nominal nutation frequency $\omega_{1,nom}$ to make them dimensionless:
\begin{eqnarray}
 {\tilde{\omega}}_0(t)                  & \equiv & \frac{ \omega_0(t)}{\omega_{1,nom}}
 \label{eq:omega0tilde} \\
  {\tilde{\omega}}_1                 & \equiv & \frac{ \omega_1}{\omega_{1,nom}}
 \label{eq:omega1tilde}
\end{eqnarray}
The nominal nutation frequency is related to the duration of the perfect refocusing pulse on resonance, $t_{180}$, by: $\omega_{1,nom} t_{180} = \pi$.
In addition, we use the normalized time $\tau$  in units of $t_E$:
\begin{equation}
 \tau                                  \equiv  \frac{ t }{t_E}
\end{equation}

\subsubsection{Effective rotation describing echo-to-echo propagator}
\begin{figure}
	   \centering
       \adjincludegraphics[width=5.0in]{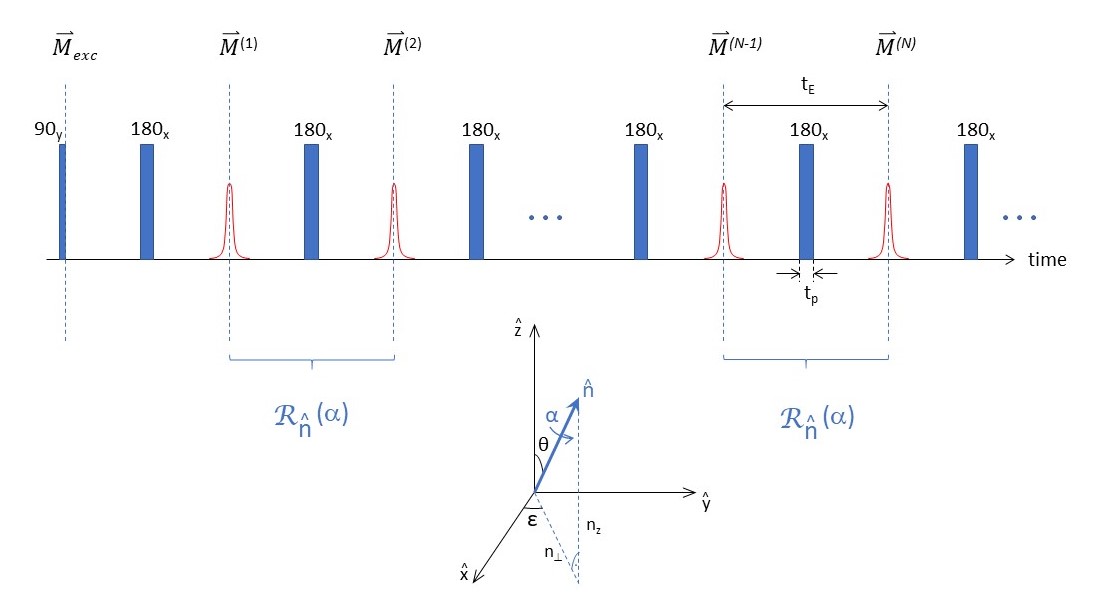}
  	 \caption[Figure 1]{CPMG sequence and indication of the echo-to-echo propagator ${\cal R}_{\hat n}(\alpha)$.}
	   \label{fig:CPMG_propagator}
\end{figure}

For the case of non-interacting spins, $\mathcal{H}_{\mathrm ave}$ can be obtained exactly and does not require a Magnus expansion.
The refocusing cycle of the basic CPMG sequence consists of three subsequent intervals (see Fig.~\ref{fig:CPMG_propagator}): a free precession interval of duration $(t_E - t_p)/2$, followed by the rf pulse of duration $t_p$, and followed by a second free precession interval of identical duration $(t_E - t_p)/2$. Each of these intervals acts as a rotation on the magnetization. The free precession intervals
correspond to rotations around $\hat z$ with an angle $\beta_1 = \omega_0 (t_E - t_p)/2$, while the rf pulse acts as a rotation around
$\left( \omega_1 \hat{x} + \omega_0 \hat{z} \right) /\sqrt{\omega_{0}^{2} +\omega_{1}^{2}}$ with an angle $\sqrt{\omega_0^2 +\omega_1^2} t_p$.
The net effect of these three subsequent rotations is a single effective rotation characterized by its rotation axis $\hat n$ and rotation angle $\alpha$. The relevant Euler parameters $\left\{ {\hat n} \sin(\alpha/2), \cos(\alpha/2) \right\}$ can be calculated efficiently using quaternions \cite{blumich85,counsell85,siminovitch97}.

The properties of the effective rotation describing the average Hamiltonian for a refocusing cycle of the CPMG sequence with static $B_0$ field (see Fig.\,\ref{fig:CPMG_propagator}) are given by \cite{hurlimann_CPMG}:
\begin{eqnarray}
n_{\perp} & = & \frac{1}{\Delta}\frac{\omega_1}{\Omega} {\sin{\beta_2}}
    \label{eq:n_perp} \\
n_z       & = & \frac{1}{\Delta}\left[{\sin{\beta_1}\cos{\beta_2}+\frac{\omega_0}{\Omega}\cos{\beta_1} \sin{\beta_2}}\right]
    \label{eq:n_z} \\
\cos\left( \frac{\alpha}{2}\right) & = & \cos{\beta_1}\cos{\beta_2}-\frac{\omega_0}{\Omega}\sin{\beta_1} \sin{\beta_2},
    \label{eq:alpha}
\end{eqnarray}
where
\begin{eqnarray}
\Omega   & = & \sqrt{\omega_{0}^{2}+ \omega_{1}^{2}}\\
\beta_1  & = &\omega_0 (t_E - t_{p})/2 \\
\beta_2  & = &\Omega t_{p}/2 \\
\Delta^2 & = &\left[ \frac{\omega_1}{\Omega}\sin{\beta_2}\right]^2 + \left[ \sin{\beta_1} \cos{\beta_2}+\frac{\omega_0}{\Omega}\cos{\beta_1}\sin{\beta_2}\right]^2.
\end{eqnarray}
The azimuthal angle $\varepsilon$ of $\hat n$ (see Fig.\,\ref{fig:CPMG_propagator}) for the static case is identical to the phase of the refocusing pulses, i.e. $\varepsilon = 0$ for $x$ and $\varepsilon = \pi/2$ for $y$-refocusing pulses.

\begin{figure}
	   \centering
        \adjincludegraphics[width=4.5in, trim={{.00
       \width} {.0\width} {.00\width} 0},clip]{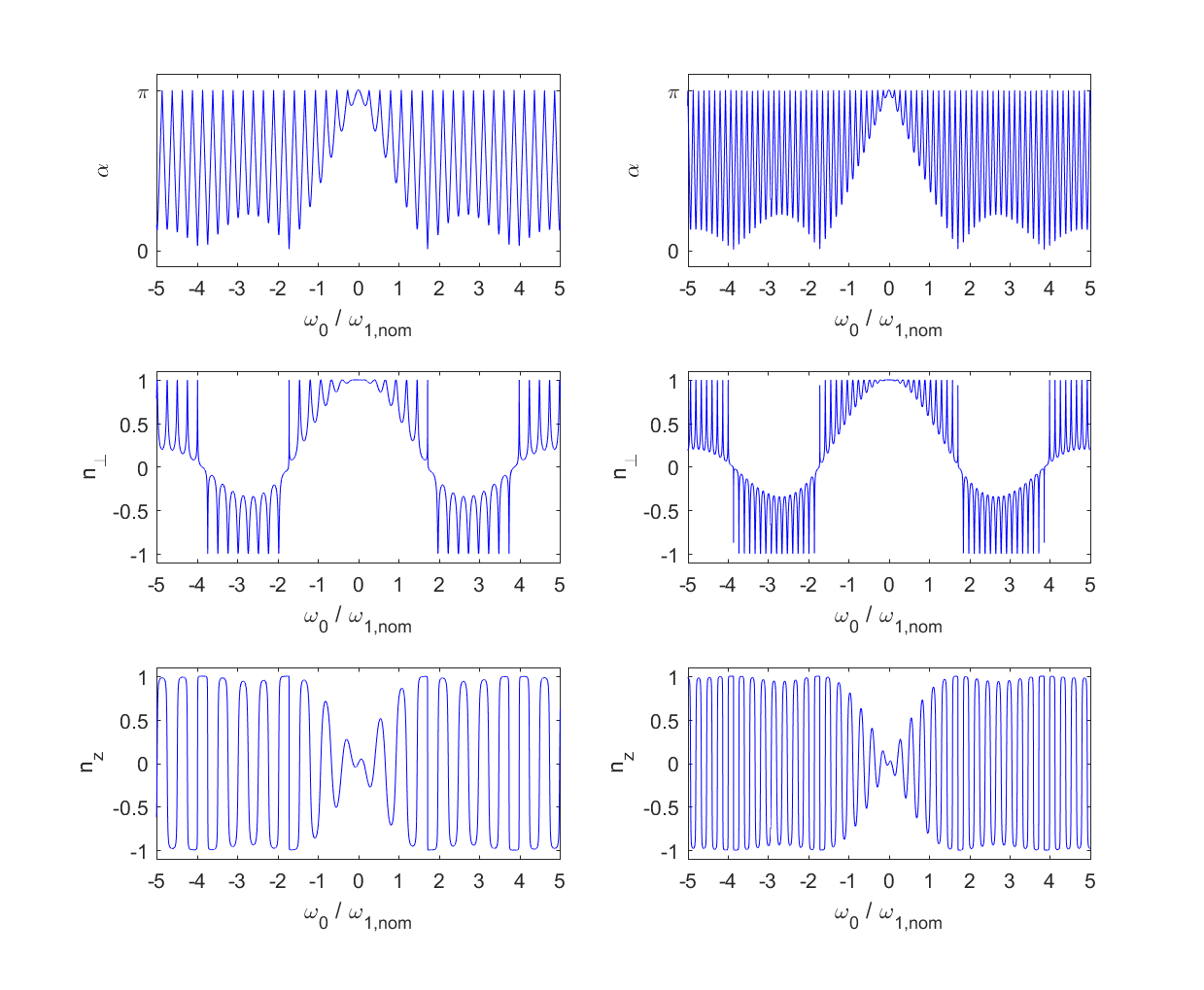}
  	 \caption[Figure 1]{Properties of the effective rotation representing the average Hamiltonian for two different echo spacings, $t_E / t_{180} = 8$ (left) and $15$ (right), respectively. The top panels show the angle $\alpha$ as a function of ${\tilde \omega}_0$ obtained from Eq.\,\eqref{eq:alpha}. Note that given the equivalence of the rotations $\left\{{\hat n},\alpha \right\}$ and $\left\{{-\hat n},\alpha \pm \pi \right\}$, it is always possible to choose $0\leq \alpha \leq \pi$. The other panels show the transverse and longitudinal component of the axis $\hat n$ obtained from Eqs.\,\eqref{eq:n_perp} and \eqref{eq:n_z}. }
	   \label{fig:alpha_nhat_vs_omega_0}
\end{figure}

Equations \eqref{eq:n_perp}-\eqref{eq:alpha} show that the angle $\alpha$ and the direction of the axis $\hat n$ describing the average Hamiltonian have a non-trivial dependence on the offset frequency $\omega_0$. This is further illustrated in Fig.\,\ref{fig:alpha_nhat_vs_omega_0} for two cases of echo spacings.

The propagator for the magnetization from the $N-1$-th to the $N$-th echo is the rotation on the Bloch sphere around the axis ${\hat n}$ with angle $\alpha$:
\begin{equation}
    {\vec M}^{(N)} = {\cal R}_{\hat n}(\alpha)\left\{ {\vec M}^{(N-1)}   \right\}.
    \label{eq:basicPropagator}
\end{equation}
In the rotating frame of the RF frequency $\omega_{rf}$, the rotation can be represented by a $3 \times 3$ matrix
with the elements $R_{jk}$ ($j,k = x,y$, or $z$) given by $
R_{jk} =  n_j n_k +(\delta_{jk} -n_j n_k) \cos \left( \alpha \right)  - \epsilon_{jkl} n_l \sin\left({\alpha} \right)$.
Here $\delta_{jk}$ is the Kronecker delta, $\epsilon_{jkl}$ is the Levi-Civita symbol, and $n_j$ is the projection of $\hat n$ onto the axis $\hat j$.

Equivalently, the rotation matrix in Eq.~\eqref{eq:basicPropagator} can be cast into the form of the SO(3) unitary transformation:
\begin{equation}
 {\cal R}_{\hat n}(\alpha)= \exp \left\{ -i \alpha {\hat n}\cdot {\vec{\mathcal{S}}}\right\}.
\end{equation}
Here $\vec{\mathcal{S}} = (\mathcal{S}_1,\mathcal{S}_2,\mathcal{S}_3)$ is a vector whose components are the three real $3 \times 3$ antisymmetric matrices $\mathcal{S}_k$ representing generators of rotations with respect to three axes in the Cartesian coordinate system. The matrix elements of $\mathcal{S}_k$ are given by $(\mathcal{S}_k)_{ij}= -i \epsilon_{ijk}$.

The $\mathcal{S}_k$ matrices also appear in the representation of un-coupled spin $S=1$ particles, which can be shown by a simple basis transformation. By rewriting the rotation matrix in the exponential form, it becomes transparent that in the continuous limit, the evolution of the magnetization from echo to echo in a CPMG sequence is equivalent to that of a spin $S=1$ particle subjected to a continuous average magnetic field, $\mathcal{\boldsymbol{B}}_{\mathit{ave}}= \frac{\alpha }{\gamma t_E} {\hat n}$, with the average Hamiltonian reading as:
\begin{equation}
\label{eq:effective-H}
\mathcal{H}_{\mathit ave} \equiv \hbar \gamma \mathcal{\boldsymbol{B}}_{\mathit{ave}} \cdot {\vec{\mathcal{S}}} = \hbar \frac{ \alpha }{t_E} {\hat n}\cdot {\vec{\mathcal{S}}} .
\end{equation}
It is, however, prudent to keep in mind that despite the mathematical similarity, our eigenstates describe the directly observable magnetization of the nuclei ensemble, unlike the quantum mechanics where wave functions are subjected to the probability interpretation.

\subsubsection{Eigenmodes of the average Hamiltonian}

It is straightforward to solve the eigenvalue problem of the average Hamiltonian of Eq.\,\eqref{eq:effective-H},
\begin{equation}
  \mathcal{H}_{\mathrm ave} {\hat v}_k = E_k {\hat v}_k.
\end{equation}
There are three eigenmodes labeled by $k = -1,0,+1$. The eigenvalues $E_k$ are given by:
\begin{equation}
 E_k = k \frac{\hbar \alpha }{t_E}.
\end{equation}
It is critical to note that the energy level associated with the $k=0$ mode is always zero. This is a general topological property of any rotation: the component of magnetization along the axis of rotation is always preserved.
\begin{figure}
	\centering
	\adjincludegraphics[width=4.5in, trim={{.00
			\width} {.0\width} {.00\width} 0},clip]{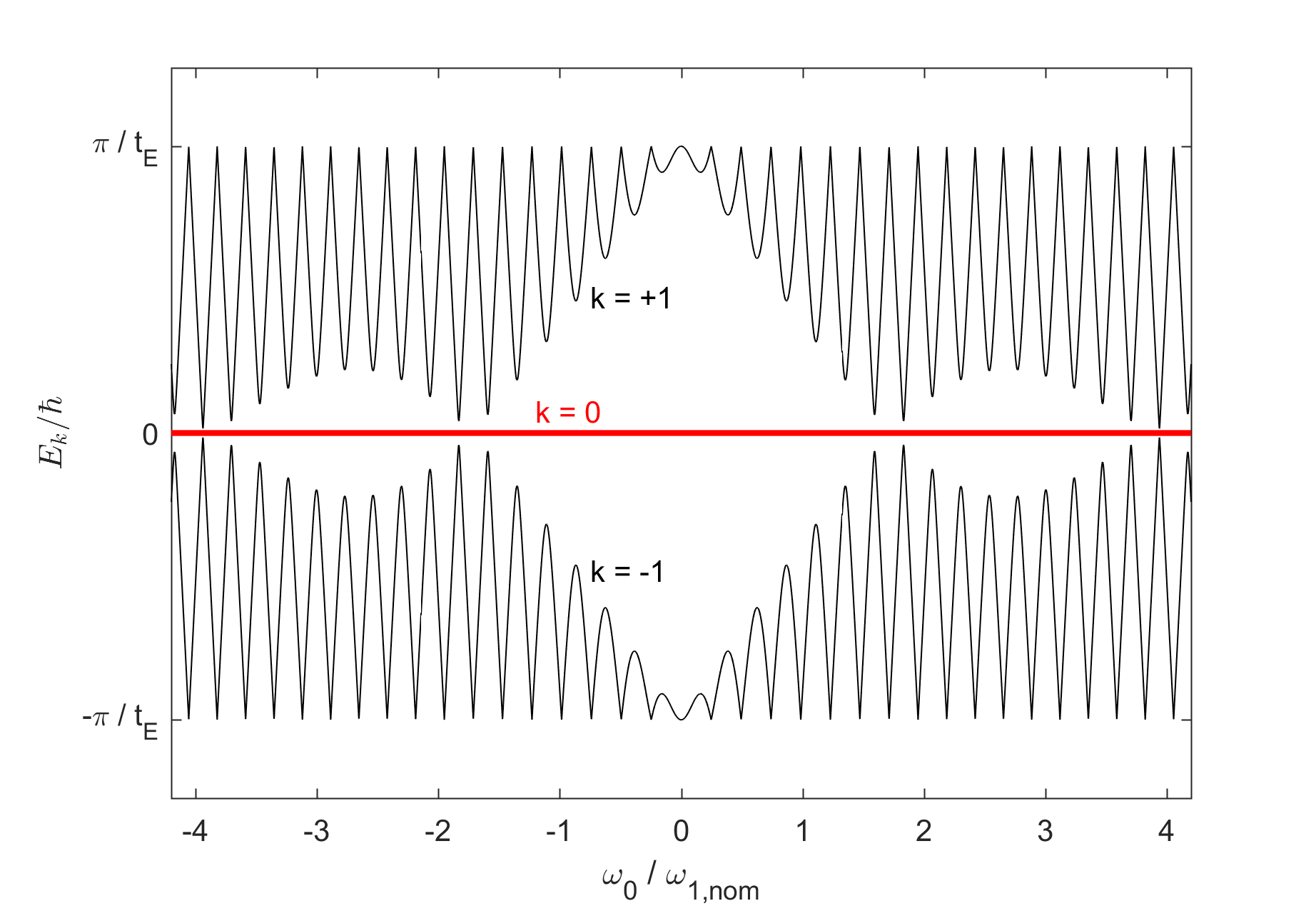}
	\caption[Figure 1]{Energy diagram for the average Hamiltonian  $\mathcal{H}_{\mathrm ave}$  as a function of the normalized offset frequency ${\tilde \omega}_0 = {\omega}_0/{\omega}_{1,nom}$ with $\omega_{1}= {\omega}_{1,nom}$. The energy level for the $k=0$ level (also referred to as CPMG level) is zero and is shown in red. Here we assumed $t_E/t_{180} = 8.1$.}
	\label{fig:alpha_vs_omega0}
\end{figure}
Figure \ref{fig:alpha_vs_omega0} shows the corresponding energy level diagram (in frequency units) as a function of offset frequency ${\tilde \omega}_0$ assuming $\omega_{1}= {\omega}_{1,nom}$. The levels for $k=\pm1$ are separated from the $k=0$ mode by $\pi / t_E$ near $\omega_0 = 0$, but are nearly degenerate with a much smaller gap at other offsets such as in the neighborhood of $\omega_0 = 1.7 \, \omega_{1,nom}$.

The three eigenvectors ${\hat v}_k$ are controlled by the direction of the axis of rotation, $\hat n$, and are independent of $\alpha$. In cylindrical coordinates of $\hat n$ (see Fig.\,\ref{fig:CPMG_propagator}), they are given by:
\begin{equation}
{\hat v}_0    =    {\hat n} =               \left( \begin{array}{c}   n_\perp \cos \varepsilon  \\
n_\perp \sin \varepsilon  \\
n_z                                       \end{array} \right);
\label{eq:v0}
\end{equation}
\begin{equation}
{\hat v}_{+1} =  \frac{-i}{\sqrt{2}} \left( \begin{array}{c}   n_z \cos \varepsilon - i \sin \varepsilon \\
n_z \sin \varepsilon + i \cos \varepsilon \\
- n_\perp                                   \end{array} \right);
\label{eq:v+1}
\end{equation}
\begin{equation}
{\hat v}_{-1} =  \frac{i} {\sqrt{2}} \left( \begin{array}{c}   n_z \cos \varepsilon + i \sin \varepsilon \\
n_z \sin \varepsilon - i \cos \varepsilon \\
- n_\perp                                   \end{array} \right).
\label{eq:v-1}
\end{equation}
Note that the eigenvector $\hat v_0$ points along the axis of rotation, $\hat n$, and $\hat v_{+1}$ and $\hat v_{-1}$ are complex conjugates of each other. As expected, the eigenvectors fulfill the orthonormal condition $ {\hat v_k}\cdot{\hat v_l}^* = \delta_{kl}$.

\subsubsection{Decomposition into CPMG and CP magnetizations}

Given the eigenmodes of the average Hamiltonian, it is straightforward to obtain an analytical expression for the magnetization at the $N$-th echo, ${\vec M}^{(N)}$ by decomposing the magnetization into the eigenmodes. For each mode, the echo-to-echo propagator is a simple phase factor $ e^{-ik\alpha}$:
\begin{equation}
 {\vec M}^{(N)} = \sum_{k=-1}^{+1} a_k \,e^{-ikN\alpha} {\hat v}_k
 \label{eq:magnetization_decomp}
\end{equation}
For the static case, the mode amplitudes $a_k$ are determined by the projection of the initial magnetization after the excitation pulse, $\vec{M}_{exc}$, onto the eigenvector ${\hat v}_k$:
\begin{equation}
 a_{k} = \vec{M}_{exc}\cdot {\hat v}_{k}^{*}.
 \label{eq:a_initial}
\end{equation}

We generally refer to the $k=0$ term as the CPMG magnetization, whereas to the sum of the $k = \pm 1$ terms as the CP magnetization.
\begin{equation}
{\vec M}^{(N)} = {\vec M}_{\mathrm CPMG}+{\vec M}_{ CP}^{(N)},
 \label{eq:M_CPMG_CP}
\end{equation}
where
\begin{eqnarray}
{\vec M}_{\mathrm CPMG}& =& a_0 {\hat v}_{0} = a_0 \hat n
  \label{eq:CPMGmagn} \\
{\vec M}_{\mathrm CP}^{(N)}&   =& a_{-1} \,e^{ikN\alpha} {\hat v}_{-1}  + a_{+1} \,e^{-ikN\alpha} {\hat v}_{+1}.
  \label{eq:CPmagn}
\end{eqnarray}
This notation is motivated by the fact that on resonance, the Carr-Purcell-Meiboom-Gill sequence \cite{meiboom58} generates ideally only the CPMG component (i.e. $a_0 = 1$, $a_{\pm 1} = 0$), whereas the original Carr-Purcell sequence \cite{carr54} generates only the CP component (i.e. $a_0 = 0$, $a_{\pm 1} = 1/\sqrt{2}$).

In general, the measured magnetization has contributions from both terms of \eqref{eq:M_CPMG_CP}. The CPMG term \eqref{eq:CPMGmagn} has a number of special properties. It is always real and its phase factor is exactly unity. This term is therefore independent of the echo number $N$. Even in inhomogeneous fields, the magnetization associated with this term remains constant from echo to echo. Note that ${\vec M}_{\mathrm CPMG}$ can be extracted from ${\vec M}^{(N)}$ by a projection onto $\hat n$:
\begin{equation}
{\vec M}_{\mathrm CPMG} = \left({\vec M}^{(N)} \cdot {\hat n} \right) {\hat n}.
\end{equation}

The $k=\pm 1$ terms in \eqref{eq:magnetization_decomp} that contribute to ${\vec M}_{CP}^{(N)}$ are in general complex. However since the $k = -1$ term is the complex conjugate of the $k=+1$ term, the CP magnetization \eqref{eq:CPmagn} is real, as any observed magnetization has to be. The phase of the CP magnetization evolves from echo to echo, but its magnitude in a homogeneous field is preserved. It can be extracted from ${\vec M}^{(N)}$ by a projection onto ${\hat v}_{+1}$ or ${\hat v}_{-1}$:
\begin{equation}
\left| {\vec M}_{\mathrm CP} \right| = \sqrt{2} \left| \left({\vec M}^{(N)} \cdot {\hat v}_{\pm 1}^{*} \right) \right| = \sqrt{2} \left|a_{\pm1} \right|.
\end{equation}
In contrast to the CPMG magnetization, the CP magnetization is strongly affected by field inhomogeneities. In inhomogeneous fields, the values for $\alpha$ show a distribution, which will lead to a distribution of
the phase factor $e^{-ikN\alpha}$ in \eqref{eq:magnetization_decomp} for $k = \pm 1$. Consequently, ${\vec M}_{\mathrm CP}^{(N)}$ in non-uniform fields will dephase with increasing echo number. This is analogous to $T_{2}^{*}$ dephasing in standard free induction measurements in inhomogeneous fields. In most practical cases, ${\vec M}_{CP}^{(N)}$ decays quickly with $N$ and  the observable magnetization ${\vec M}^{(N)}$ is well approximated by ${\vec M}_{\mathrm CPMG}$  at high echo numbers. In grossly inhomogeneous fields as studied in \cite{hurlimann_CPMG} this occurs already after the second echo.

\subsection{Time-dependent $B_0(t)$: generalized Landau-Zener problem}
\label{sec:LandauZener}

In the previous section, we treated the time-dependence of the rf field $B_1$ in the original Hamiltonian \eqref{eq:Hamiltonian_orig} using the average Hamiltonian approach for an arbitrary, but constant value of $B_0$. We now take into account the time dependence of ${\vec B}_0(t)$. As mentioned in the introduction, we assume here that $d{\tilde \omega}_0 / d\tau \ll 1$.

When $B_0$ and $B_1$ are time-dependent, the propagator for each cycle can still be expressed in terms of a rotation operator.  For each refocusing cycle $j$, we can still define an average Hamiltonian in the rotating frame associated with $\omega_{rf}$ with its corresponding eigenvectors and phase factors. However, the eigenvectors and eigenvalues now generally change from one refocusing cycle to the next.
Formally, the evolution of magnetization for the dynamic case can be written as:
\begin{equation}
{\vec M}^{(N)} = {\cal R}_{\hat n^{(N)}}(\alpha^{(N)})\left\{ {\cal R}_{\hat n^{(N-1)}}(\alpha^{(N-1)}) \ldots
\left\{ {\cal R}_{\hat n^{(1)}}(\alpha^{(1)})
\left\{ {\vec M}_{exc} \right\} \right\}
\ldots \right\}.
\label{eq:M_general}
\end{equation}
As in Eq.~\eqref{eq:M_CPMG_CP} for the static case, we can still decompose the magnetization into corresponding CPMG and CP components at each cycle in the dynamic case. However, transitions between the CPMG and CP components now become possible from cycle to cycle, as the eigenmodes change.

The formal expression of magnetization evolution in Eq.~\eqref{eq:M_general} can be cast into the form in terms of the average Hamiltonians ${\mathcal{H}}_{\mathrm ave}\left(\alpha^{(j)},n^{(j)} \right)$ as:
\begin{equation}
{\vec M}^{(N)} = \mathcal{T} \left\{ \prod_{j=1}^N \exp\left[-i \frac{\mathcal{H}_{\mathrm ave}\left(\alpha^{(j)},{\hat n}^{(j)} \right)}{\hbar} t_E\right] \right\} {\vec M}_{exc},
\label{eq:M_general_H}
\end{equation}
where $\mathcal{T} \left\{ \dots \right\}$ is the time ordered operator. The average Hamiltonian $\mathcal{H}_{\mathrm ave}\left(\alpha^{(j)},{\hat n}^{(j)} \right)$ is still defined by Eq.~\eqref{eq:effective-H}, but with the rotation angle $\alpha^{(j)}$ and the rotation axis ${\hat n}^{(j)}$ now specified at each cycle.

To lowest order in $d{\tilde \omega}_0 / d\tau $, $\alpha^{(j)}$ and ${\hat n}^{(j)}$ can be obtained from the expressions \eqref{eq:n_perp} to \eqref{eq:alpha} at the instantaneous values of $\omega_0$ and $\omega_1$ at the $j$-th cycle. $B_0$ field variation during the refocusing cycle $j$ leads to a small correction to the azimuthal angle $\varepsilon$ that describes the direction of $n_{\perp}^{(j)}$. To first order in $d{\tilde \omega}_0 / d\tau $, the correction is given by:
\begin{equation}
\delta\varepsilon^{(j)} =  \frac{\pi}{8} \frac{t_E}{t_{180}} \frac{ d{\tilde{\omega}}_0^{(j)}}{d \tau}.
\label{eq:epsilon}
\end{equation}
There is no first order correction  to the angle $\alpha^{(j)}$ from its instantaneous value.

\subsection{Slow fluctuation: adiabatic regime}

Now, the problem at hand is a particular case of evolution under a time-dependent Hamiltonian. When the variations of magnetic fields and hence of the corresponding Hamiltonian are sufficiently slow, we expect that the evolution of the time dependent eigenstates follows the instantaneous eigenstates of the Hamiltonian up to the dynamic phase and a geometric phase~\cite{born_fock_1928}. As a result, the general expression for the adiabatic evolution of the magnetization at the $N$-th echo, Eq.~\eqref{eq:M_general},
simplifies to:
\begin{equation}
{\vec M}^{(N)}_{adia} = \sum_{k = -1}^{+1} \left(\vec{M}_{\mathrm exc} \cdot {\hat v}_{k}^{(1)*}\right) \exp \left\{-ik \sum_{j=1}^{N}\alpha^{(j)} +i \sum_{j=1}^{N-1} \Gamma_k^{(j)}\right\} {\hat v}_{k}^{(N)}.
\label{eq:Magn_adia}
\end{equation}
Here, the term $\left(\vec{M}_{exc} \cdot {\hat v}_{k}^{(1)*}\right) = a_{k}^{(1)}$ is the initial amplitude of mode $k$. It depends on the initial magnetization after the excitation pulse, $\vec{M}_{exc}$ and the eigenvector of the first refocusing cycle, ${\hat v}_{k}^{(1)}$. The term $\exp\left\{-ik\sum_{j=1}^{N}\alpha^{(j)}\right\}$ accounts for the accumulated dynamic phase during all the refocusing cycles, while the term $\exp\left\{i\sum_{j=1}^{N-1}\Gamma_k^{(j)}\right\}$ describes the accumulated geometric phase with $\Gamma_k^{(j)}$ given by \cite{berry84}:
\begin{equation}
\label{eq:geometric_phase}
\Gamma_k^{(j)}= -\Im \left\{\ln\left[ {\hat v}_{k}^{(j)*} \cdot {\hat v}_{k}^{(j+1)}  \right] \right\},
\end{equation}
where $\Im \left\{ \ldots \right\}$ takes the imaginary part of the argument.
In \eqref{eq:Magn_adia}, the last term $ {\hat v}_{k}^{(N)}$ is the eigenvector of the final refocusing cycle $N$. In the adiabatic limit, there are no transitions between CPMG and CP components.

For the particular interest of the CPMG ($k=0$) component, the dynamic phase vanishes throughout the time evolution as $E_0$ is exactly zero for this component. Critically, the geometric phase also vanishes exactly as ${\hat v}_{0}^{(j)} =  {\hat n}(t_j)$ is real for each refocusing cycle. We hence obtain the key result:
\begin{equation}
{\vec M}^{(N)}_{CPMG,adia} = \left(\vec{M}_{exc} \cdot {\hat n}(t_1)\right) {\hat n}(t_N)= a_{0}^{(1)} {\hat n}(t_N).
\label{eq:CPMG_adia}
\end{equation}
This result shows that in the adiabatic regime, the magnetization of the CPMG component only depends on the value of $B_0$ and $B_1$ at the beginning of the sequence and
at the echo time of interest. It does not depend on the field variation between these points. In particular for periodic variations when the field returns to its original value, the magnetization of the CPMG component will also return to its initial value. In contrast, the magnetization of the CP component retains a path dependence through both the dynamic phase, $\phi_{dyn} = {\sum_{j=1}^{N}\alpha^{(j)}}$, and the geometric phase, $\phi_{geo} = {\sum_{j=1}^{N-1}\Gamma^{(j)}_{\pm 1}}$ in Eq.~\eqref{eq:Magn_adia}.

The results of \eqref{eq:Magn_adia} and \eqref{eq:geometric_phase} are expressed in the rotating frame of the rf frequency $\omega_{rf}$. In this case, the appearance of the geometric phase can be thought of as a manifestation of the generalized Coriolis effect as the direction of the eigenfunction changes with time. Alternatively, Garwood {\it et al.}\cite{garwood2001} analyzed adiabatic pulses using a doubly rotating frame of reference where this effect manifests itself instead as an additional fictitious field that is proportional to the rate of change of the average field.

\subsection{Adiabatic condition and beyond the adiabatic regime}

To take full advantage of the Hamiltonian language, we will now promote the magnetization evolution in Eq.~\eqref{eq:M_general_H} from discrete operations into the continuous formalism, such that
\begin{equation}
{\vec M}(t) =  \exp\left[- \frac{i}{\hbar}  \int_0^{t} \mathcal{H}_{\mathit ave}\left(t'\right) dt' \right] {\vec M}_{exc},
\label{eq:M_general_H_con}
\end{equation}
where the time-dependent Hamiltonian is given by
\begin{equation}
\mathcal{H}_{\mathit ave}(t) \equiv \hbar \gamma \mathcal{\boldsymbol{B}}_{{\mathit{ave}}}(t) \cdot {\vec{\mathcal{S}}} = \hbar \frac{\alpha(t) }{t_E} {\hat n(t)}\cdot {\vec{\mathcal{S}}}.
\label{eq:H_con}
\end{equation}
As we will discuss later, this continuous evolution formalism allows us to understand qualitatively (or quantitatively when the continuous limit is well justified) the transition between CPMG and CP modes in the dynamic cases.

The validity of the adiabatic evolution given by Eqs.~\eqref{eq:Magn_adia} and ~\eqref{eq:CPMG_adia} depends on how fast the magnetic field is changing. Following the arguments developed for standard fast passage experiments \cite{baum85,tannus97}, the adiabatic condition requires that the direction of the average field (or here $\hat n$) has to change with a rate slower than the instantaneous nutation frequency. As long as  the variations of magnetic fields do not accelerate or decelerate drastically, we can safely ignore the change of $\hat{n}(t)$ in the azimuthal direction and the rate of change of the direction of $\mathcal{\boldsymbol{B}}_{\mathit{ave}}(t)$ is given by $d \theta /dt$. Here, $\theta$  (see Fig.\,\ref{fig:CPMG_propagator}) can be obtained by the relationships $n_{\perp} = \sin(\theta)$ and $n_z = \cos(\theta)$ from Eqs.~\eqref{eq:n_perp} and~\eqref{eq:n_z}. The effective nutation frequency is given by $\alpha / t_E$. This results in the adiabatic condition:
\begin{equation}
 \left| \frac{d \theta}{dt} \right| \ll \left| \frac{\alpha}{t_E} \right|.
  \label{eq:adiabaticity_condition}
\end{equation}

It is also useful to consider the current problem as an example of a Landau-Zener type of problem~\cite{Landau1932,Zener1932,Stueckelberg1932,Majorana1932}. As the offset frequency $\omega_0$ increases from on-resonance $\omega_0=0$ to a larger absolute value, the effective energy gap shown in Fig.~\ref{fig:alpha_vs_omega0} oscillates and reduces to near zero at certain values where a strong "diabatic" transition between the CPMG and CP modes can occur.
To quantify transitions between the CPMG and CP modes induced by the variations of magnetic fields, it is most convenient to use the continuous formalism for the magnetization evolution given in Eqs.~\eqref{eq:M_general_H_con} and~\eqref{eq:H_con}. In the continuous limit, the adiabatic evolution of the magnetization reads as:
\begin{equation}
{\vec M}_{adia} (t) = \sum_{k = -1}^{+1} \left(\vec{M}_{exc} \cdot {\hat v}_{k}^{*}(0)\right) \exp \left\{ -\frac{i}{\hbar} \int_0^t E_k(t') dt' +i \Gamma_k(t) \right\} {\hat v}_{k}(t),
\label{eq:Magn_adia_con}
\end{equation}
with the geometric phase $\Gamma_k(t)$ given by \cite{berry84}:
\begin{equation}
\label{eq:geometric_phase_con}
\Gamma_k (t)= i \int_{0}^t dt' {\hat v}_{k}^{*}(t') \cdot \partial_{t'}{\hat v}_{k}(t').
\end{equation}
Similar to the discrete case, the term $\left(\vec{M}_{exc} \cdot {\hat v}_{k}^{*}(0)\right)$ is the initial amplitude $a_{k}(0)$ of mode $k$ at $t=0$, the integral of $E_k$ gives the dynamical phases, and $ {\hat v}_{k}(t)$ is the eigenvector at the instantaneous time $t$.

As noted, Eq.~\eqref{eq:Magn_adia_con} is only an approximate solution to the original effective Hamiltonian in Eq.~\eqref{eq:H_con} with the effective field $\mathcal{\boldsymbol{B}}_{\mathit{ave}}(t)$ in the adiabatic limit. However following the argument and derivation by M. V. Berry in Ref.~\cite{Berry2009}, it is an exact solution to a related problem with a Hamiltonian of the same functional form but with a modified effective field $\mathcal{\boldsymbol{B}}_{\mathit{mod}}$ as:
\begin{equation}
\gamma \mathcal{\boldsymbol{B}}_{\mathit{mod}}(t)=\gamma \mathcal{\boldsymbol{B}}^{\ }_{\mathit{ave}}(t) + \hat{n}(t)\times \partial_{t} \hat{n}(t).
\label{eq:mod_B_eff}
\end{equation}
The modified field $\mathcal{\boldsymbol{B}}_{\mathit{mod}}(t)$ consists of the sum of the original effective field $\mathcal{\boldsymbol{B}}^{\ }_{\mathit{ave}}(t)$ and the additional fictitious field  proportional to $\hat{n}(t)\times \partial_{t} \hat{n}(t)$. This fictitious field  is governed by the tangential variation of the rotation direction, $\hat{n}(t)$ (or equivalently the direction of the effective field $\mathcal{\boldsymbol{B}}^{\ }_{\mathit{ave}}(t)$).

Some quick observations are worth mentioning here. First, a comparison of the magnitude of the fictitious field to that of the original effective field allows us to evaluate the validity of the adiabatic approximation. The magnitude of the fictitious field is $\left| \frac{d \theta}{dt} \right|$, while the magnitude of the original field is $\alpha / t_E$. We thus recover the adiabatic condition stated in \eqref{eq:adiabaticity_condition}. Second and less directly, an inverted version of Eq.~\eqref{eq:mod_B_eff} can be used for the evaluation of transition between the CPMG and CP when the continuous limit is well justified, which we will discuss in Appendix.

\subsection{Critical ramp rate $\nu_{0,crit}$ and adiabaticity parameter $\cal A$}

\begin{figure}
	   \centering
        \adjincludegraphics[width=4.0in, trim={{.00
       \width} {.0\width} {.00\width} 0},clip]{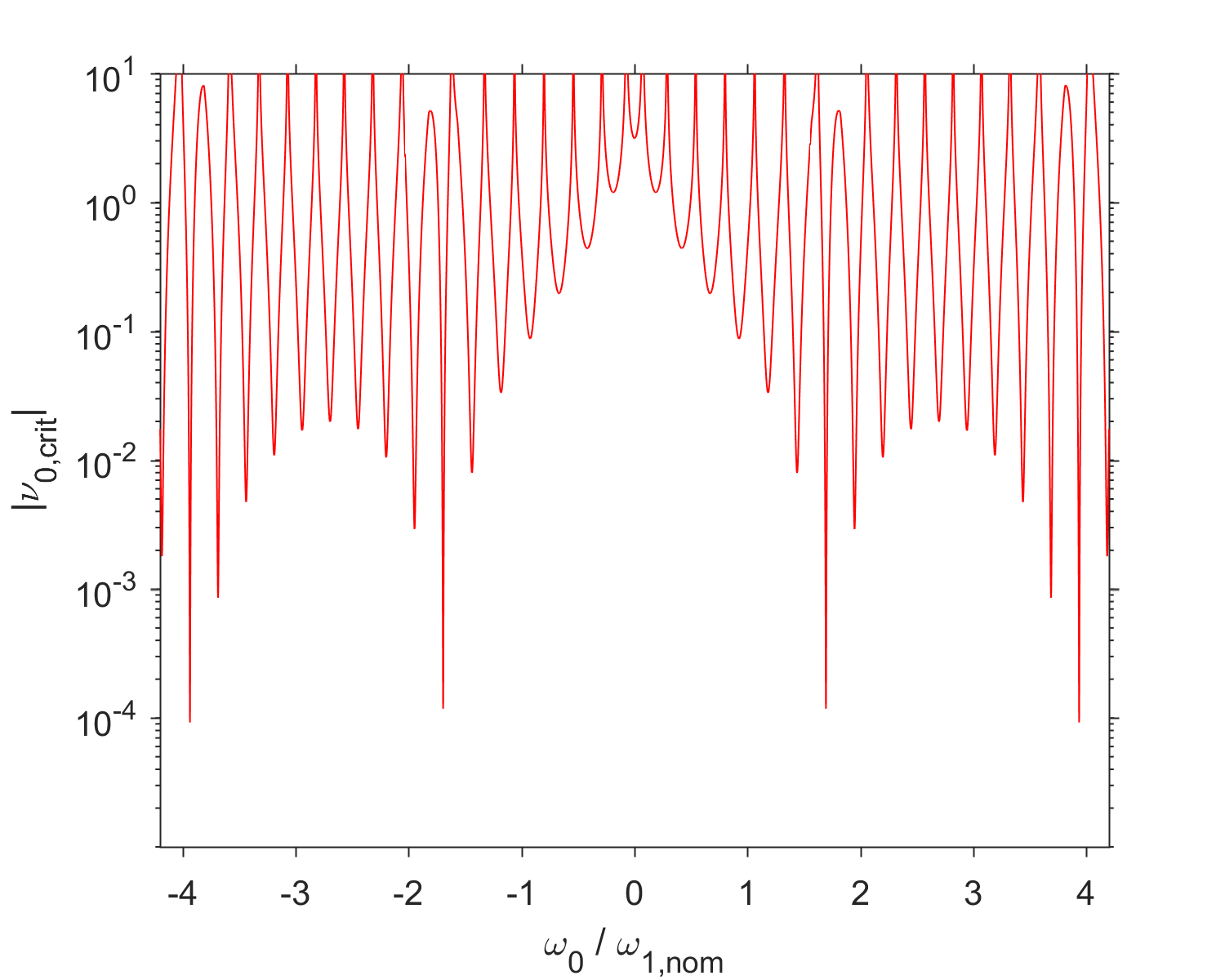}
  	 \caption[Figure 1]{
Critical ramp rate $\nu_{0,crit}$ versus $\tilde{\omega}_0$ calculated from Eq.\,\eqref{eq:nu_0_definition}. Here we assumed $\omega_1 = \omega_{1,nom}$ and $t_E / t_{180}= 8.1$
  }
	   \label{fig:nu_0crit}
\end{figure}

The adiabatic condition in \eqref{eq:adiabaticity_condition} is expressed in terms of the quantities $\left\{ \theta,\alpha \right\}$ that refer to the direction and amplitude of the effective field $\mathcal{\boldsymbol{B}}^{\ }_{\mathit{ave}}(t)$.
To relate the adiabatic condition more directly to the rate of change of the applied magnetic field $B_0$, we rewrite the change of the direction of $\hat n$ in terms of the rate of change of ${\tilde \omega}_0$ by:
\begin{equation}
\frac{d \theta}{dt} = \frac{d \theta}{d \omega_0}\frac{d \omega_0}{dt} =
     \frac{1}{t_E}\frac{d \theta}{d {\tilde \omega}_0}\frac{d {\tilde \omega}_0 }{d\tau}.
\end{equation}
Here we have assumed that $B_1$ is constant in time. It is useful to introduce the critical ramp rate of $\nu_{0,crit}$ defined by:
\begin{equation}
 \nu_{0,crit} \equiv \frac{\alpha}{{d\theta}/{d {\tilde \omega}_0}},
 \label{eq:nu_0_definition}
\end{equation}
so that the adiabatic condition \eqref{eq:adiabaticity_condition} becomes
\begin{equation}
\frac{d {\tilde \omega}_0 }{d\tau} \ll \nu_{0,crit}.
 \label{eq:adiabatic_condition_critical_vel}
\end{equation}
The critical ramp rate is a dimensionless quantity that depends only on the static rotation properties of the refocusing cycle and can be calculated from the expressions Eqs.\,\eqref{eq:n_perp}, \eqref{eq:n_z}, and \eqref{eq:alpha}. Figure~\ref{fig:nu_0crit} shows the dependence of $\nu_{0,crit}$ on the offset frequency ${\tilde \omega}_0$.
The critical ramp rate in Fig.\,\ref{fig:nu_0crit} shows a high sensitivity on ${\tilde \omega}_0$ with pronounced minima near ${\tilde \omega}_0 = \pm 1.7$ and $\pm 3.9$. This is consistent with the energy level diagram of Fig.\,\ref{fig:alpha_vs_omega0} that shows that at these offsets, the energy splitting between the CPMG and CP levels nearly vanishes and transitions are easier to induce.

It is useful to introduce the adiabaticity parameter $\cal A$ defined as the ratio of the critical ramp rate $\nu_{0,crit}$ and the dimensionless experimental ramp rate ${d {\tilde \omega}_0 }/{d\tau}$:
\begin{equation}
{\cal A} \equiv \left| \frac{ \nu_{0,crit}}{\frac{d {\tilde \omega}_0 }{d\tau}} \right|
\label{eq:adiabaticity}
\end{equation}
The adiabatic condition \eqref{eq:adiabaticity_condition} now becomes simply ${\cal A} \gg 1$.

\subsection{Extension to time variable $B_0$ and $B_1$ fields}
\label{sec:B1motion}

\begin{figure}
	   \centering
        \adjincludegraphics[width=5.0in, trim={{.00
       \width} {.0\width} {.00\width} 0},clip]{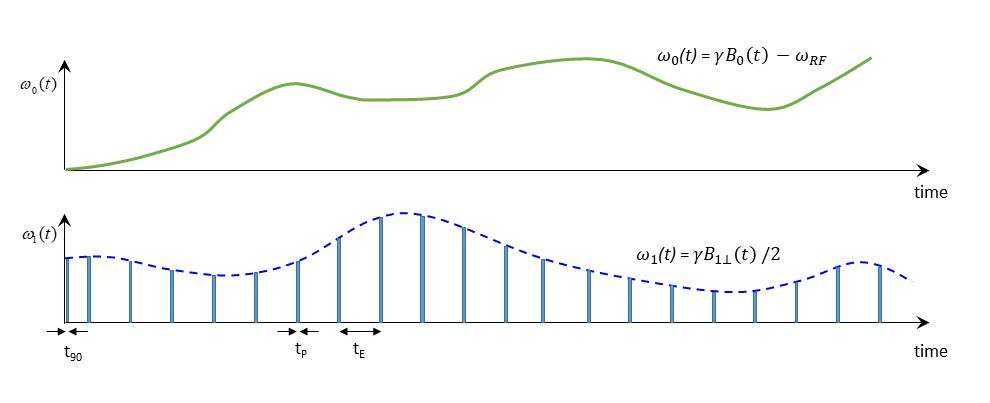}
  	 \caption[Figure 1]{General case of CPMG sequence with fluctuating $B_0$ and $B_1$ fields, leading to time dependence of $\omega_0$ and $\omega_1$. }
	   \label{fig:B0_B1_motion}
\end{figure}

To obtain a better understanding of the origin of the minima in $\nu_{0,crit}$, is it instructive to consider the more general case when both  $B_0$ and $B_1$ fields are time-dependent, illustrated in Fig.\,\ref{fig:B0_B1_motion}.
In this treatment, we neglect the temporal variation of $B_1$ during the duration of an individual rf pulse, but take into account the change of $B_1$ from one pulse to the next. In this case, the evolution is still described by Eq.~\eqref{eq:M_general} with the average Hamiltonian of the form Eq.~\eqref{eq:H_con} with the direction and amplitude of the effective field $\left\{ \theta, \alpha \right\}$ given by the expressions Eq.~\eqref{eq:n_perp} to Eq.~\eqref{eq:alpha} and Eq.~\eqref{eq:epsilon}. These quantities now depend not only on the instantaneous value of $\tilde{\omega}_0(t)$ but also on $\tilde{\omega}_1(t)$, but otherwise the treatment carries through. Similarly, the expressions for the adiabatic regime Eq.~\eqref{eq:Magn_adia} and Eq.~\eqref{eq:CPMG_adia} are still applicable when both $B_0$ and $B_1$ are fluctuating, as is the general form for the adiabatic condition Eq.~\eqref{eq:adiabaticity_condition}.

However, the expression for the adiabaticity parameter $\cal A$ in \eqref{eq:adiabaticity} has to be generalized to take into account the fluctuations of $B_1$.
Given that the nutation frequency of the pulses, $\omega_1 = \gamma B_{1,\perp}/2$, is now time dependent, it is essential to take this into account.
In the general case, the temporal derivative of the angle $\theta$ has to be expanded into temporal derivatives of $\tilde{\omega}_0$ and $\tilde{\omega}_1$:
\begin{equation}
\frac{d \theta\left(\tilde{\omega}_0(\tau),\tilde{\omega}_1(\tau)\right)}{d \tau} =
\frac{d \theta}{d\tilde{\omega}_0}\frac{d\tilde{\omega}_0}{d \tau}+
\frac{d \theta}{d\tilde{\omega}_1}\frac{d\tilde{\omega}_1}{d \tau}
\end{equation}
It is useful to introduce now an additional critical ramp rate for $B_1$, $\nu_{1,crit}$ that sets the scale of admissible $B_1$ variations within the adiabatic regime:
\begin{equation}
  \nu_{1,crit} \equiv \frac{\alpha}{{d\theta}/{d {\tilde \omega}_1}}
 \label{eq:nu_1_definition}
\end{equation}

The generalized adiabaticity parameter now becomes:
\begin{equation}
\frac{1}{\cal A} =
\left| \frac{1}{\nu_{0,crit}}\frac{d {\tilde \omega}_0 }{d\tau} + \frac{1}{\nu_{1,crit}}\frac{d {\tilde \omega}_1 }{d\tau} \right|.
\label{eq:adiabaticCondition_B0B1}
\end{equation}
With this generalized definition of the adiabaticity parameter, the adiabatic condition remains ${\cal A} \gg 1$.

\begin{figure}
	   \centering
        \adjincludegraphics[width=5.5in, trim={{.00
       \width} {.0\width} {.00\width} 0},clip]{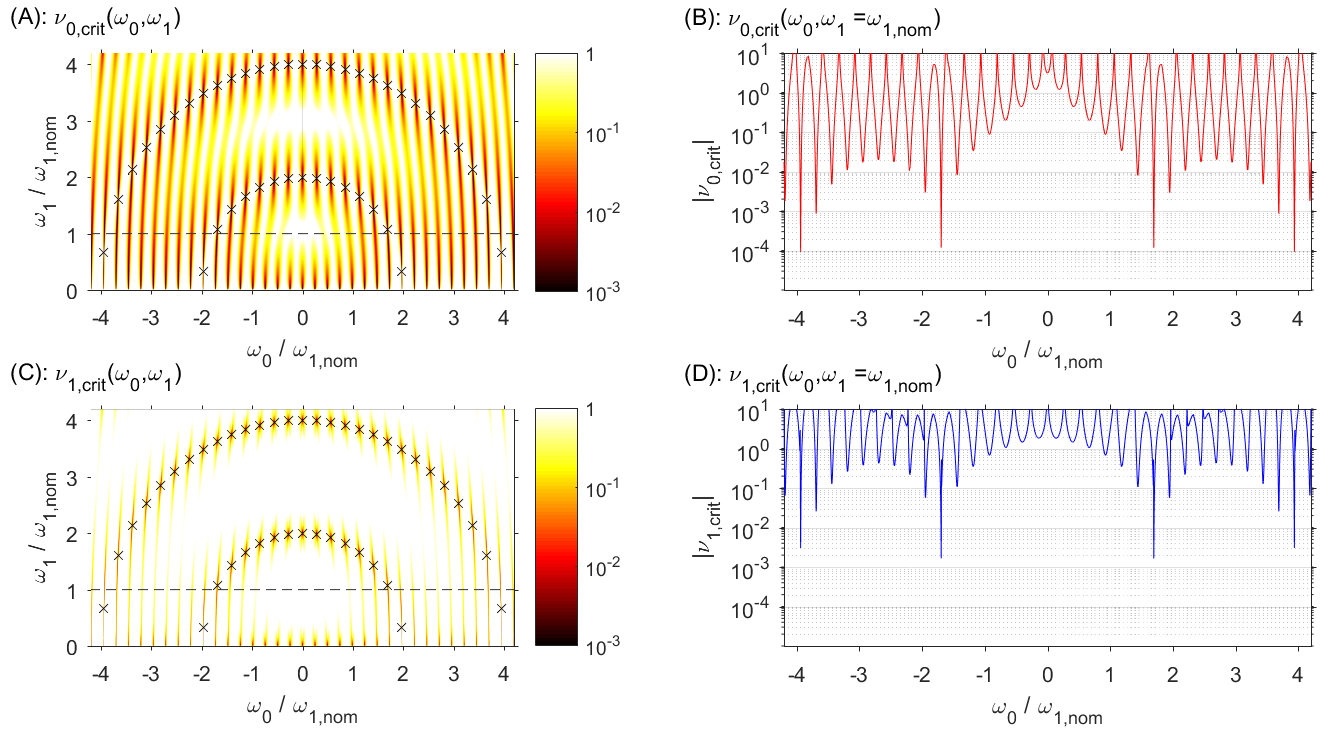}
  	 \caption[Figure 1]{Critical ramp rate $\nu_{0,crit}$ (top) and $\nu_{1,crit}$ (bottom) versus scaled $B_0$ offsets, ${\tilde \omega}_0$, and scaled $B_1$ amplitude, ${\tilde \omega}_1$. The right panels, show cross sections for $\omega_1 = \omega_{1,nom}$ that is indicated as a dashed line on the left panels. Also shown on the left panels by crosses are the locations of the singularities Eq.\,\eqref{eq:singularities} where the propagator becomes the unity operator and the critical velocities vanish. }
	   \label{fig:nu_2d_2ndversion}
\end{figure}

Figure \ref{fig:nu_2d_2ndversion} shows the two critical ramp rates $\nu_{0,crit}$ and $\nu_{1,crit}$ as a function of ${\tilde \omega}_0$ and ${\tilde \omega}_1$. In addition, cross sections for $\omega_1 = \omega_{1,nom}$ are shown on the right panels.

The critical ramp rates  $\nu_{0,crit}$  and $\nu_{1,crit}$  vanish at a discrete values of $({\tilde \omega}_0,{\tilde \omega}_1)$ that are indicated by crosses in Fig.\,\ref{fig:nu_2d_2ndversion}. They correspond to singular points where the propagator becomes exactly the unity operator. These singular points occur at:
\begin{equation}
\left( \tilde{\omega}_0,\tilde{\omega}_1\right)_{\mathrm{singular},l,m} = \left(
\pm\frac{2(l-m)}{\frac{t_E}{t_{180}}-1},
\sqrt{(2l)^2-\left(  \frac{2(l-m)}{\frac{t_E}{t_{180}}-1}  \right)^2}
\right)
\label{eq:singularities}
\end{equation}
where $l$ and $m$ are integrals with $l \leq m < l \frac{t_E}{t_{180}}$.
For a unity operator, the nutation angle $\alpha = 0$. As a consequence, the CPMG and CP levels become degenerate leading to transitions between CPMG and CP modes.

When the $B_0$ - $B_1$ fluctuation corresponds to a trajectory in the $({\tilde \omega}_0,{\tilde \omega}_1)$ plane that intersects any of these singular points, the adiabaticity parameter $\cal A$ momentarily becomes 0 and the overall spin dynamics cannot be adiabatic. On trajectories that do not intersect but get close to these points, $\cal A$ will drop, but for sufficiently slow passage around these points it is possible to prevent transitions between the CPMG and CP levels.
The singular points for a particular value of $l$ lie on a circle of radius $2l$ in the $\tilde{\omega}_0 - \tilde{\omega}_1$ plane. When the $B_1$ field is fixed at its nominal value (i.e. $\tilde{\omega}_1 = 1$), the trajectory of any $B_0$ fluctuation corresponds to a straight horizontal line in the  $({\tilde \omega}_0,{\tilde \omega}_1)$ plane that intersects these circles at
$\tilde{\omega}_0 = \pm \sqrt{4l^2 - 1} = \pm (1.73, 3.87, 5.92,\ldots)$. This explains why the pronounced minima in $\nu_{crit,0}$ and $\nu_{crit,1}$ are observed in Fig.\,\ref{fig:nu_0crit} and Fig.\,\ref{fig:nu_2d_2ndversion} occur near these offset frequencies.

\section{Comparison of theory with numerical simulations}
\label{sec:Results}

To test the validity of the analytical results, we compare the analytical results with numerical simulations. We used the scalable fast C++ simulation code developed by Colm Ryan and first described in \cite{duan2017}. The algorithm is based on the integration of the Bloch equations and allows a general pulse sequence with arbitrary and possibly time-dependent ${\vec B}_0(t)$ and ${\vec B}_1(t)$ fields. The simulations account for the effects of the rf pulses under possible off-resonance conditions and precession between the pulses in ${\vec B}_0(t)$. For the current results, we set the $1/T_1$ and $1/T_2$ relaxation rates to zero.

\subsection{Linear ramp}
We first present results for the simple case of a linear field ramp. It is assumed that at the start of the CPMG sequence, the $B_0$ field fulfills the Larmor condition. The deviation of the instantaneous Larmor frequency from the RF frequency is then of the form: ${\omega}_0(t) /{\omega_{1,\mathrm{nom}}}= \frac{d{\tilde \omega}_0}{d \tau} t/{t_E} $. In practice, such a linear ramp can be caused by temporal fluctuations in the magnet or by motion of the sample relative to a magnetic field  characterized by a gradient $g$. In that case with a relative velocity $v$, the dimensionless ramp rate is given by $\frac{d{\tilde \omega}_0}{d \tau} = \gamma  g v t_E / {\omega_{1,\mathrm{nom}}}$.

\subsubsection{Simulation of magnetization}
\begin{figure}
	   \centering
        \adjincludegraphics[width=4.5in, trim={{.00
       \width} {.0\width} {.00\width} 0},clip]{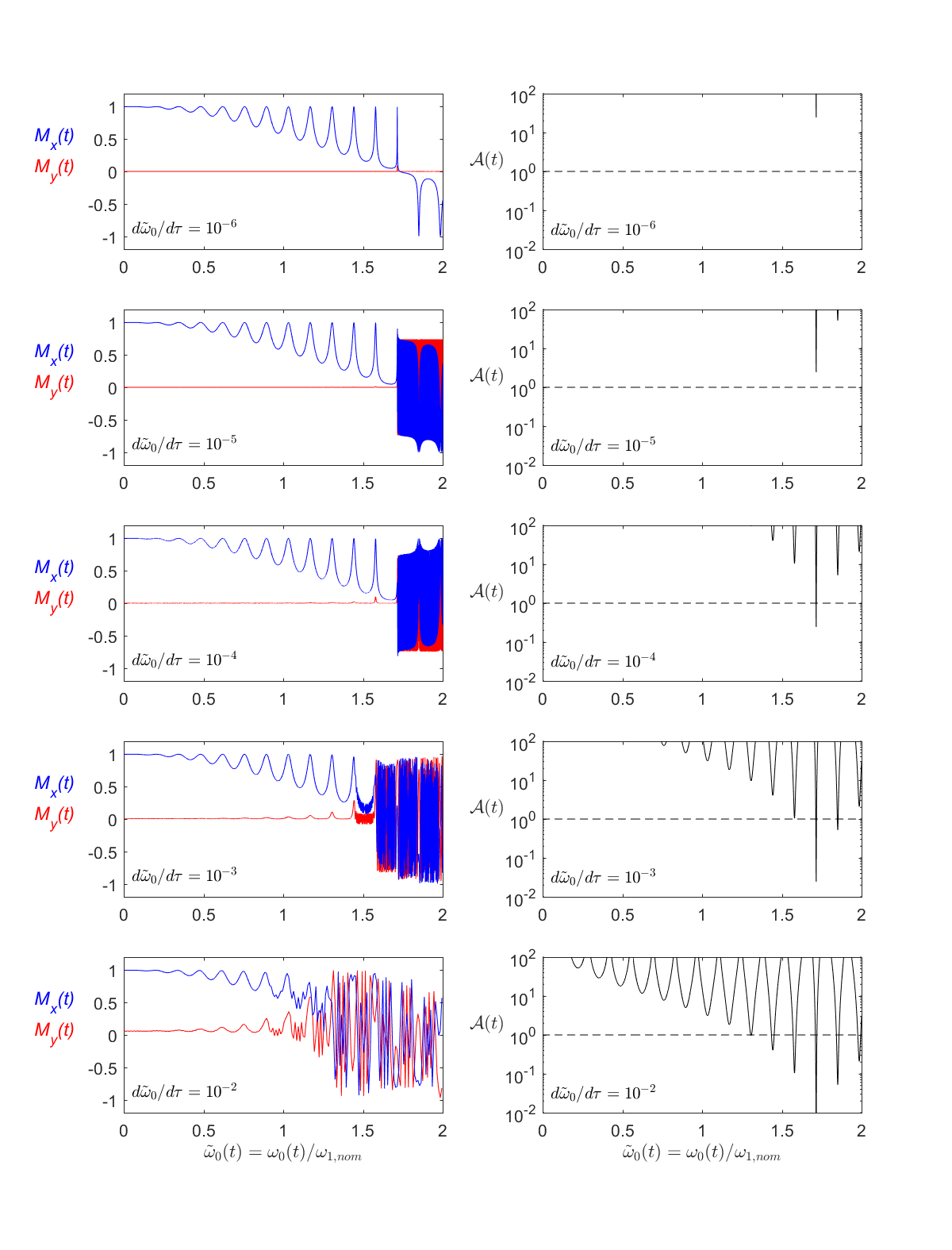}
  	 \caption[Figure 1]{Results for linear $B_0$ ramp starting from resonance with different ramp rates $\frac{d{\tilde \omega}_0}{d \tau}$ as indicated, $t_E / t_{180} = 15$, and $\omega_1 = \omega_{1,nom}$. The left panels show the simulation results of the transverse magnetization (blue: in-phase, red: out-of-phase) as a function of the instantaneous offset frequency. The right panels show the calculated adiabaticity parameters $\cal A$ from Eq.\,\eqref{eq:adiabaticity} as a function of the instantaneous offset frequency. The dashed line indicates ${\cal A} = 1$. }
	   \label{fig:lin_motion_magnetization_5ex}
\end{figure}

Fig.\,\ref{fig:lin_motion_magnetization_5ex} shows the results for five different ramp rates.
For small offset frequencies, ${\cal A} \gg 1$ for all ramp rates. This implies that the spin dynamics starts in the adiabatic regime for all cases. Furthermore, since the direction of the magnetization after the initial $90^\circ$ pulse, $M_{exc}$, coincides with the eigenvector for the $k=0$ (CPMG) mode, only this mode is initially occupied. Eq.\,\eqref{eq:CPMG_adia} then predicts that the magnetization follows ${\hat n}$:  ${\vec M}(t) = {\hat n}\left( {\tilde \omega}_0(t)\right)$. In this regime, the magnetization is determined by the instantaneous offset frequency ${\tilde \omega}_0(t)$ and does not otherwise depend on the ramp rate. The simulation results in Fig.\,\ref{fig:lin_motion_magnetization_5ex} at small to moderate offset frequencies are in full agreement with this prediction. The oscillations in $M_x$ correspond to the variations in the direction of $\hat n$ within the $\hat{x}-\hat{z}$ plane, as shown in Fig.\,\ref{fig:alpha_nhat_vs_omega_0}. The magnetization is effectively spin-locked to ${\hat n }(t)$, or equivalently to $\mathcal{\boldsymbol{B}}^{\ }_{\mathit{ave}}(t)$. At the highest ramp rate, a small $\hat y$ component in $\vec M$ becomes noticeable even at small offset frequencies. This is in agreement with Eq.\,\eqref{eq:epsilon} that shows that $\hat n$ is slightly tilted from the  $\hat{x}-\hat{z}$ plane. At the highest ramp rate of $\frac{d{\tilde \omega}_0}{d \tau} = 10^{-2}$, Eq.~\eqref{eq:epsilon} predicts $\delta \varepsilon =  3.4^\circ$, which is in good agreement with the phase of the simulated magnetization at small values of ${\tilde \omega}_0$ in Fig.\,\ref{fig:lin_motion_magnetization_5ex}.

At the lowest ramp rate, the calculated adiabaticity parameter fulfills the adiabaticity condition ${\cal A} \gg 1$ at all offset frequencies considered. Therefore the magnetization is predicted to follow Eq.\,\eqref{eq:CPMG_adia} and to stay in the CPMG mode over the entire frequency range without any transition to the CP mode. This is confirmed in the simulations. At higher ramp rates, there are regions of offset frequencies where the adiabatic condition is not fulfilled anymore. In these regions, the simulation results show indeed transitions between the CPMG and CP modes. The first onset is readily apparent in the simulations of Fig.\,\ref{fig:lin_motion_magnetization_5ex} by the appearance of fast echo fluctuation and the appearance of large out-of-phase components. These features indicate a sizeable population of the CP modes. As predicted from the adiabaticity parameters displayed on the right panels, the appearance occurs at smaller offset frequencies when the ramp rate is increased.

\subsubsection{Mode amplitudes}
We can make this comparison between simulation and theory more quantitative by decomposing the magnetization obtained from the numerical simulation into the CPMG and CP components. Knowing the field fluctuation ${\tilde \omega}_0(t)$ allows us to calculate the eigenvectors ${\hat v}_k(t)$ from Eq.\,\eqref{eq:v0} to Eq.\,\eqref{eq:v-1} for any time. The mode amplitudes $a_k(t)$ can then be extracted from the simulated magnetization data by $a_k(t) = {\vec M}(t)\cdot {\hat v}_{k}^{*}(t)$.
The amplitude of the CPMG component corresponds to $a_0(t)$. The CP component consists of the $k = \pm 1$ modes and the corresponding amplitudes are in general fluctuating rapidly due to phase factors even in the adiabatic regime. However the magnitude of the CP component, given by $\left| a_{CP}(t)\right| = \sqrt{2}\left| a_{\pm 1}(t)\right|$ is expected to be constant in the adiabatic regime.

\begin{figure}
	   \centering
        \adjincludegraphics[width=4.5in, trim={{.00
       \width} {.0\width} {.00\width} 0},clip]{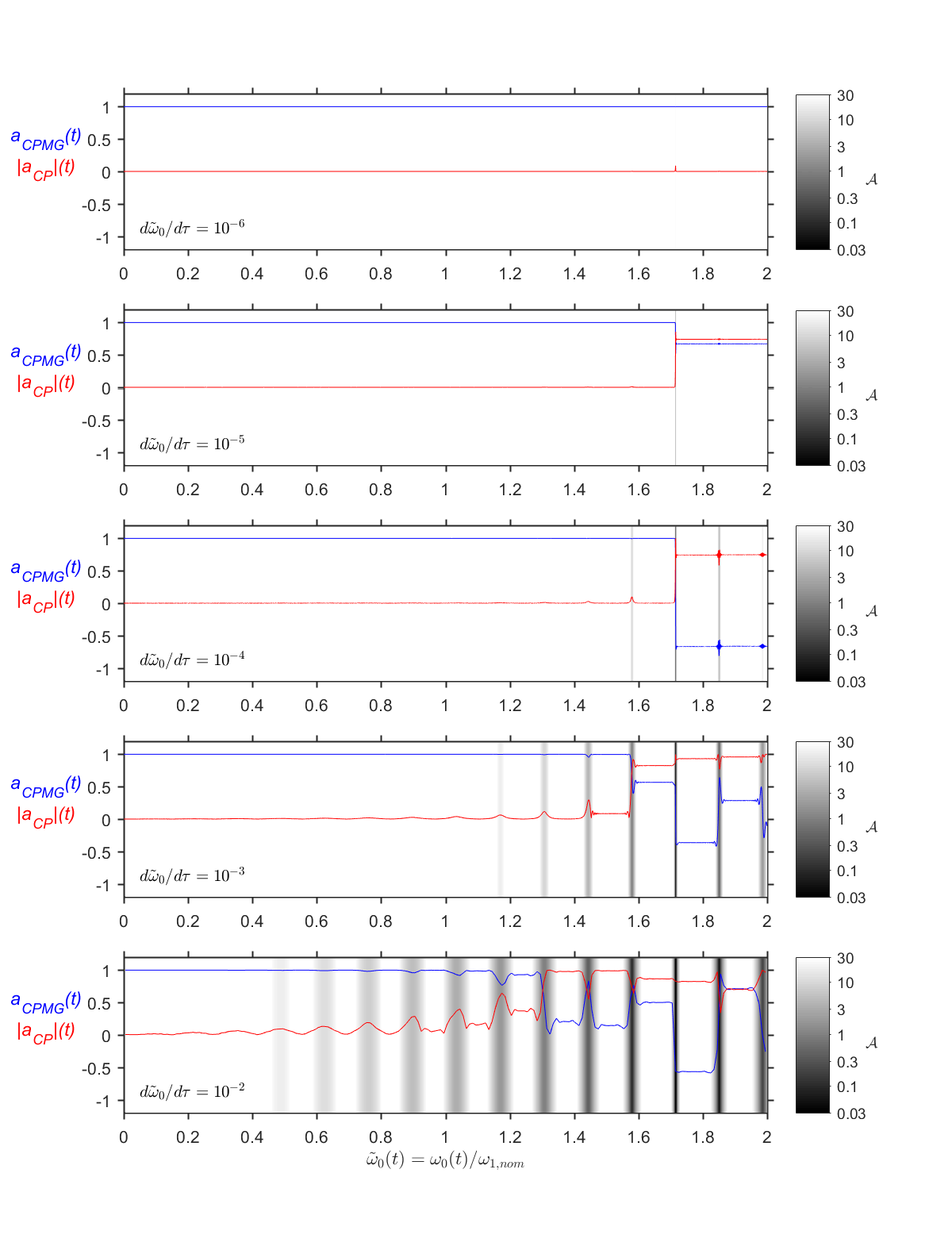}
  	 \caption[Figure 1]{CPMG amplitudes (blue) and CP magnitudes (red) extracted from the simulation results shown in Fig.\,\ref{fig:lin_motion_magnetization_5ex}. The instantaneous adiabaticity parameter ${\cal A}$ is shown in gray scale. The white regions indicate adiabatic regions where the CPMG amplitudes and CP magnitudes stay constant. }
	   \label{fig:lin_motion_amp_5ex}
\end{figure}
Figure \ref{fig:lin_motion_amp_5ex} displays the values of the CPMG amplitude and the CP magnitude extracted from the simulation as described. Superimposed to these results in gray scale are values of the adiabaticity parameter $\cal A$ (same data as in the right panel of Fig.\,\ref{fig:lin_motion_magnetization_5ex}). The adiabatic and non-adiabatic region can be easily identified as white and dark sections, respectively. In the adiabatic regions, the extracted amplitude of the CPMG mode and the magnitude of the CP mode stay indeed constant and only change in the non-adiabatic regions. Until the first non-adiabatic region is reached, $a_{CPMG} = 1$. Figure \ref{fig:lin_motion_amp_5ex} demonstrates that the adiabaticity parameter $\cal A$ can accurately predict the onset of
transitions between the CPMG and CP mode. Furthermore, it confirms the prediction that
the number and extent of these non-adiabatic regions increases at higher ramp rates.

\subsubsection{First order corrections to analytical results}

\begin{figure}
	\centering
	\adjincludegraphics[width=4.5in, trim={{.00
			\width} {.0\width} {.00\width} 0},clip]{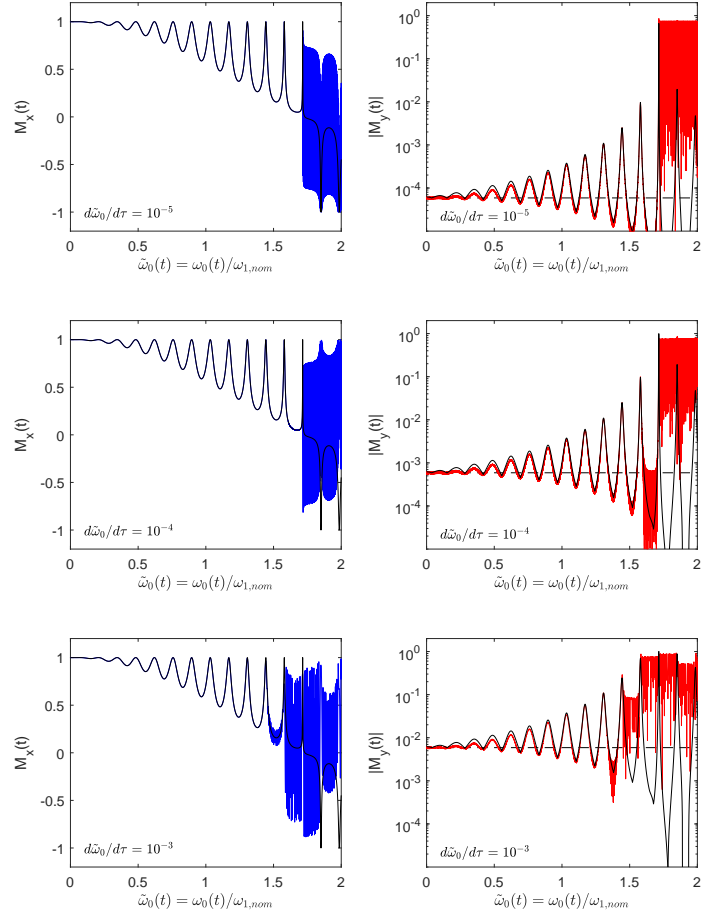}
	\caption[Figure 1]{Comparison of numerical simulations and first order analytical results. The magnetizations $M_x$ (left panels) and $|M_y|$ (right panels) are shown versus normalized offset frequency for linear $B_0$ ramps starting from resonance  with different ramp rates $\frac{d{\tilde \omega}_0}{d \tau}$ as indicated, $t_E / t_{180} = 15$, and $\omega_1 = \omega_{1,nom}$. The blue and red curves are results from numerical simulation, while the black curves are the first order prediction from Eq.~\eqref{eq:Mx-My-first-order}. The black dashed line in the right panels shows the value of $sin(\delta \epsilon) $.}
	\label{fig:first-order}
\end{figure}

At modest ramp rates, it is interesting to observe the appearance and subsequent disappearance of $M_y$ components in Fig.\,\ref{fig:lin_motion_magnetization_5ex} and of the related CP amplitudes in Fig.\,\ref{fig:lin_motion_amp_5ex}.
In the strictly adiabatic regime ($1/{\cal A}\rightarrow 0$), the evolution of the magnetization is controlled by
Eq.\,\eqref{eq:CPMG_adia} and is expected to be:
\begin{eqnarray}
M_x^{(0)}(t)   &  = & \cos(\delta \epsilon)  n_{\perp}(t) .
\label{eq:Mx-zero-order}\\
|M_y^{(0)}(t)| &  = & \sin (\delta \epsilon) n_{\perp}(t)
\label{eq:My-zero-order}
\end{eqnarray}
In this adiabatic limit, the magnitude of $M_y$ component is predicted to be bounded by  $\sin (\delta \epsilon)$. In the linear displays of Fig.\,\ref{fig:lin_motion_magnetization_5ex} and Fig.\,\ref{fig:lin_motion_amp_5ex}, deviations are clearly visible in regions with ${1/\cal A} \sim 0.1$. When $M_y$ is displayed on a logarithmic scale (see Fig.\,\ref{fig:first-order}),
it is evident that there are systematic deviations even for much smaller values of ${1/\cal A}$.
These deviations can be understood based on the inverted version of Eq.~\eqref{eq:mod_B_eff} (c.f. Eq.~\eqref{eq:inverted_B_eff} in \ref{app:B}). A perturbation calculation to first order in $1/\mathcal{A}$ yields:
\begin{equation}
\label{eq:Mx-My-first-order}
\begin{split}
M_x^{(1)}(t) & = \frac{1}{\sqrt{1+ 1/\mathcal{A}(t)^2}} \left[ \cos (\delta \epsilon) n_{\perp}(t) - 1/\mathcal{A}(t) \sin (\delta \epsilon) \right],
\\
|M_y^{(1)}(t)| & = \frac{1}{\sqrt{1+ 1/\mathcal{A}(t)^2}} \left[ \sin (\delta \epsilon) n_{\perp}(t) +  1/\mathcal{A}(t) \cos (\delta \epsilon)  \right].
\end{split}
\end{equation}
Here, the first terms come from the in-phase component that follows the adiabatic evolution. On the other hand, the second terms represent the out-of-phase component normal to the plane defined by $\hat{z}$ and $\hat{n}$ vectors, which drive the magnetization evolution away from the adiabatic trajectory. Comparisons of the simulation results and the first order perturbation predictions shown in Fig.~\ref{fig:first-order} demonstrate that the evolution of magnetizations during a linear ramping process is essentially captured by the straightforward first order perturbation calculations until a sizable transition occurs and the perturbation theory becomes invalid.
At lower levels, we note discrepancies of $|M_{y}|$ between the simulation results and the perturbation predictions that are of the order of the dimensionless ramp rate $d{\tilde \omega}_0/d\tau$. These deviations can be attributed to the inherent discrete nature of the CPMG pulse sequence that is neglected in the average Hamiltonian picture.

It is feasible to extend the perturbation expansion to higher orders as described for instance in reference \cite{deschamps2008}. As discussed in the appendix, Eq.~\eqref{eq:inverted_B_eff} describes the problem exactly in the continuous limit and can be solved numerically to give an exact answer.

\subsubsection{Dependence of mode amplitudes on ramp rate}

\begin{figure}
	   \centering
        \adjincludegraphics[width=4.0in, trim={{.00
       \width} {.0\width} {.00\width} 0},clip]{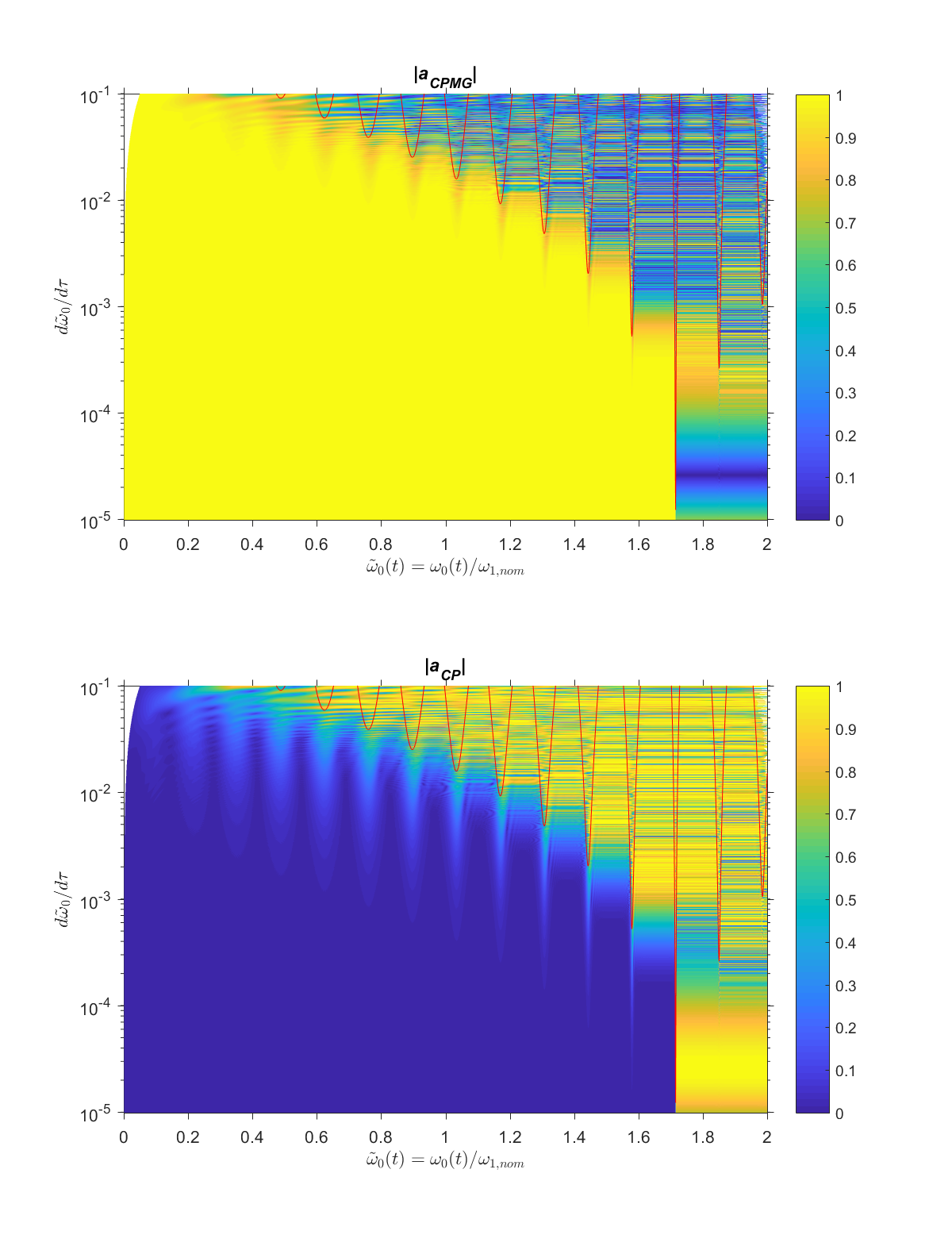}
  	 \caption[Figure 9]{Dependence of the eigenmode levels, extracted from the numerical simulations of the magnetization evolution during the CPMG, on field offset and ramp rate for linear field ramp. Magnitudes of the CPMG (top) and CP modes (bottom) are displayed versus normalized offset frequency ${\tilde \omega}_0$ and ramp rate $\frac{d {\tilde \omega}_0}{d \tau}$ for linear field ramp starting from resonance. The red line indicates the location where ${\cal A} = 2.$ The simulations where performed for  $t_E / t_{180} = 15$. }
	   \label{fig:lin_motion_2d}
\end{figure}

The results presented so far have confirmed that the adiabaticity parameter $\cal A$ is able to predict the locations and the widths of the regions where transitions between the CPMG and CP modes can occur. However, the knowledge of $\cal A$ is not sufficient do predict the change in the mode amplitude. As illustrated in Fig.\,\ref{fig:lin_motion_2d}, these changes are highly sensitive to the ramp rate.

Whenever a non-adiabatic region (delineated approximately by the red line) is encountered, the mode amplitude changes and assumes a new constant level in the next adiabatic region. A very small change in ramp rate can lead to a completely different level. The traversing of the non-adiabatic regions always lasts many refocusing cycle. Even small changes in the timing of the pulses can accumulate to large changes in the resulting magnetization. As a consequence, the robust prediction of the magnetization after traversing a non-adiabatic region is in general very challenging as it will require a very detailed knowledge of the field fluctuations that is often not available.

\subsection{Harmonic Motion}

\begin{figure}
	   \centering
        \adjincludegraphics[width=4.5in, trim={{.00
       \width} {.0\width} {.00\width} 0},clip]{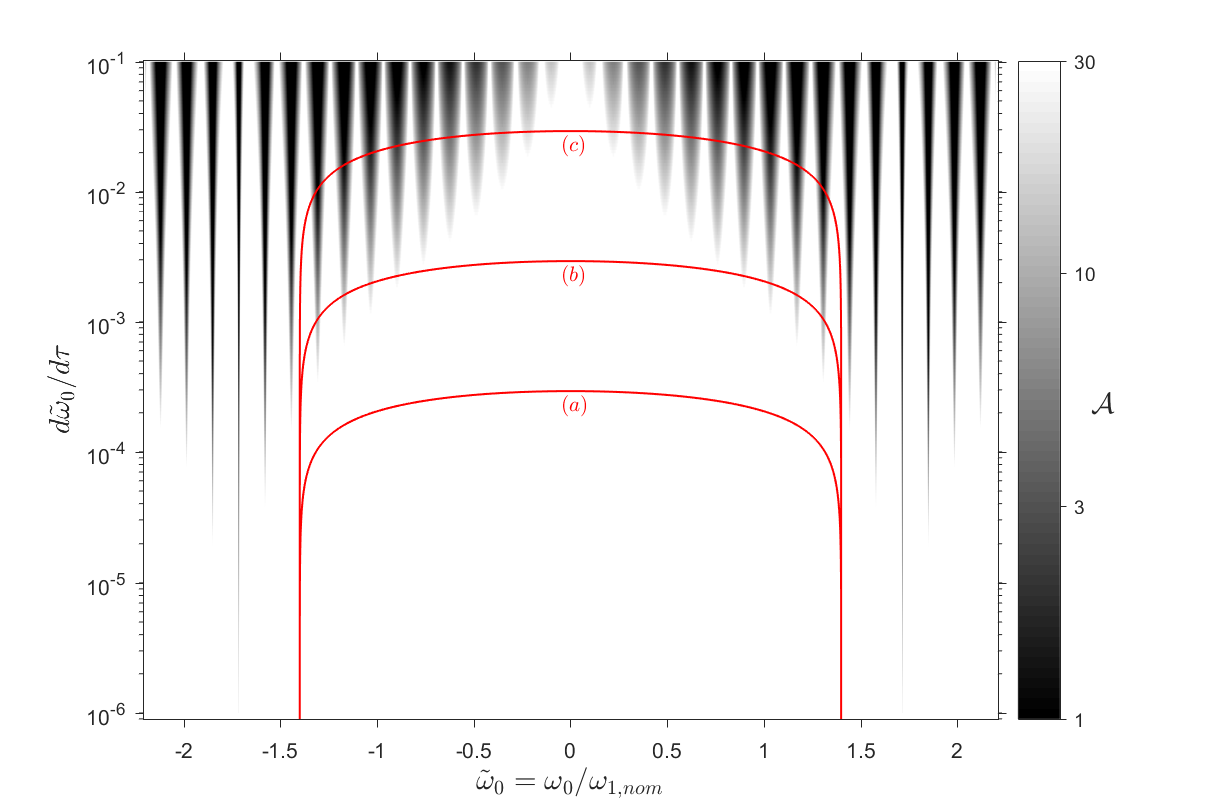}
  	 \caption[Figure 1]{Paths of harmonic fluctuations (red) superimposed over the adiabaticity parameter $\cal A$ (shown in gray scale) in the ${\tilde \omega}_0 - \frac{d {\tilde \omega}_0}{d \tau}$ parameter space. The three harmonic motions are of the form $\omega_0 (t) = \Delta \omega_0 sin(2 \pi t/T)$ with an amplitude of $\Delta \omega_0 = 1.4 \omega_{1,nom}$ and periods $T$ of (a) $T =  3 \times 10^4 t_E$, (b)  $T =  3002 t_E$  and (c) $T =  300.2 t_E$. In all cases, $t_E / t_{180} = 15$. }
	   \label{fig:harmonicMotion_Phasediagram}
\end{figure}

The examples presented so far were characterized by linear field fluctuations with a constant ramp rate $\frac{d {\tilde \omega}_0}{d \tau}$. Here we demonstrate that the analysis developed in section \ref{sec:Theory} applies to more general fluctuations with varying rates. The key quantity is the instantaneous adiabaticity parameter ${\cal A}(t)$ that controls whether the spin dynamics is in the adiabatic or non-adiabatic regime. This parameter is determined by the instantaneous offset frequency  ${\tilde \omega}_0(t)$ and its instantaneous rate of change, $\frac{d {\tilde \omega}_0}{d \tau}(t)$. We can therefore characterize a particular $B_0$ fluctuation by its path in the ${\tilde \omega}_0 - \frac{d {\tilde \omega}_0}{d \tau}$ parameter-space to determine the evolution of ${\cal A}(t)$. This approach is illustrated in Fig.\,\ref{fig:harmonicMotion_Phasediagram} for harmonic motions.

The white areas in Fig.\,\ref{fig:harmonicMotion_Phasediagram} indicate adiabatic regions, whereas the dark areas indicate non-adiabatic regions. If the path of the fluctuation is entirely confined to the white area (for example the path labelled $(a)$ in Fig.\,\ref{fig:harmonicMotion_Phasediagram}), then the spin dynamics is essentially adiabatic and Eq.\,\eqref{eq:Magn_adia} can be applied to lowest order. In this case, it is straightforward to predict the response of the system. The amplitude of the CPMG mode, $a_{CPMG}$, and the magnitude of the CP component, $|a_{CP}|$, stay constant. In contrast, when the path intersects non-adiabatic regions (e.g. path $(c)$ in  Fig.\,\ref{fig:harmonicMotion_Phasediagram}), transitions between the CPMG and CP modes occur and the behavior becomes very sensitivity to the details of the path. We can accurately predict the onset of the chaotic behavior, but the prediction of the magnetization at later times becomes problematic as the response becomes extremely sensitive to the external parameters.

\begin{figure}
	   \centering
        \adjincludegraphics[width=4.5in, trim={{.00
       \width} {.0\width} {.00\width} 0},clip]{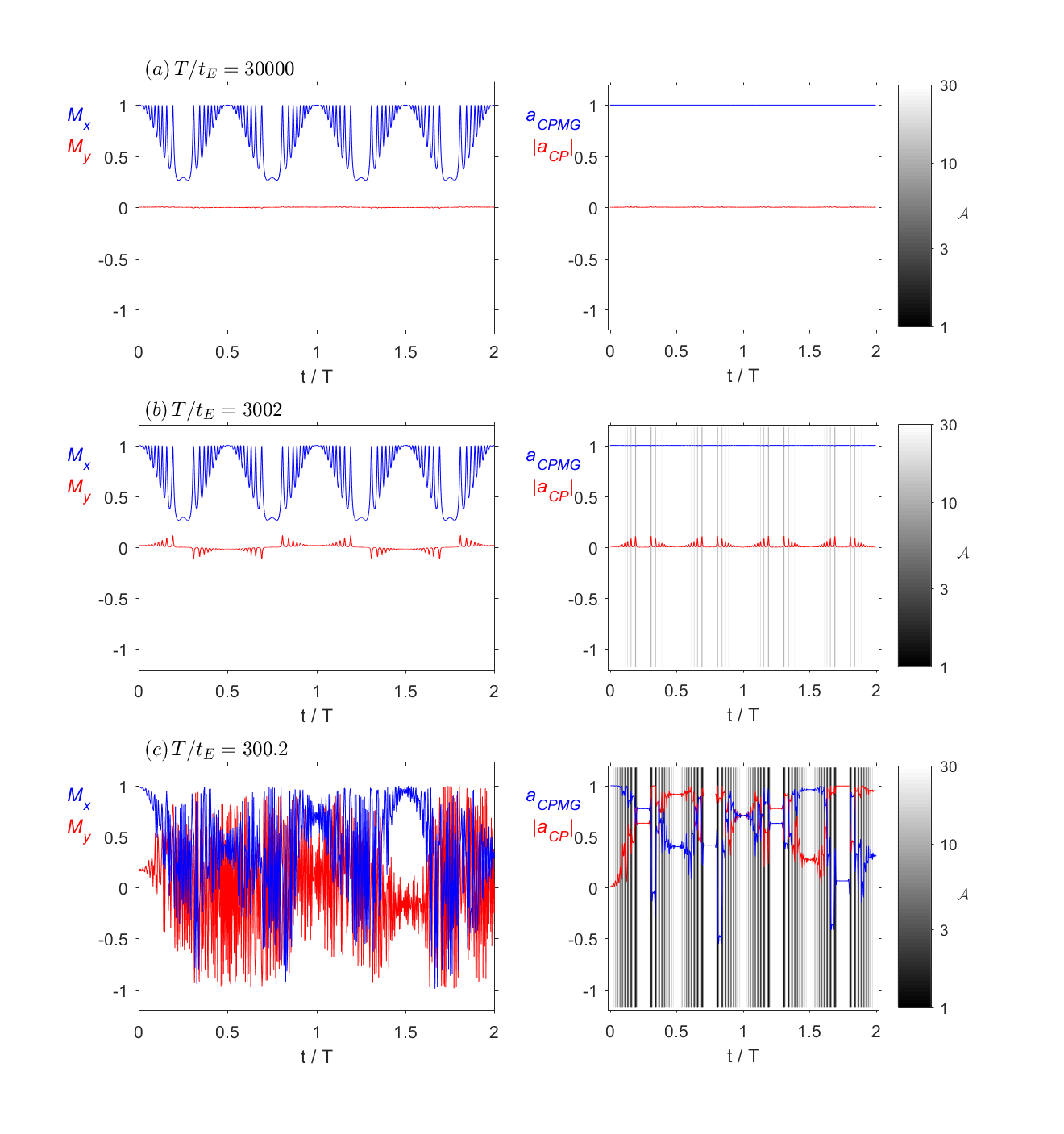}
  	 \caption[Figure 1]{Results for harmonic field variations $B_0(t)$ with an amplitude of $1.4 \, B_1$ starting from resonance. The left panel shows the resulting transverse magnetization over 2 periods (blue in-phase, red out-of-phase). The period $T$ is varied from $T =  3 \times 10^4 t_E$ (top) to $T =  3002 t_E$ (middle) and $T =  300.2 t_E$ (bottom).
  The right panels show the extracted CPMG amplitudes (blue) and CP magnitudes (red). The adiabaticity parameters $\cal A$ are shown in gray scale. For the three values of $T$, the minimum adiabaticity parameters are 91, 9.1, and 0.93, respectively. In all cases, $t_E / t_{180} = 15$. }
	   \label{fig:harmonicMotion}
\end{figure}

Figure~\ref{fig:harmonicMotion} shows the time dependence of the magnetization and mode amplitudes obtained by numerical simulation for the three harmonic motions indicated in Fig.\,\ref{fig:harmonicMotion_Phasediagram}. The results for the longest period (path $a$ in Fig.\,\ref{fig:harmonicMotion_Phasediagram}) show a purely in-phase magnetization that is periodic in time and synchronized with the fluctuation. The mode decomposition shows that the signal only contains the CPMG component. The magnetization is therefore perfectly spin-locked to the instantaneous direction of $\hat n$ and the evolution fully adiabatic. This simulation result is in full agreement with the simple analysis of its path in Fig.\,\ref{fig:harmonicMotion_Phasediagram}. The path $a$ is completely confined to the adiabatic region with a minimum value of ${\cal A}(t)$ of $91 \gg 1$.

In contrast, the response for the shortest period $T$ shown in Fig.\,\ref{fig:harmonicMotion}(c) shows only a small region of adiabatic behavior similar to that in Fig.\,\ref{fig:harmonicMotion}(a), but then becomes 'noisy' and chaotic. The response is not periodic in time anymore. Again this is anticipated by the simple analysis of its path in Fig.\,\ref{fig:harmonicMotion_Phasediagram}. The minimum value of the adiabaticity parameter for path $c$ is less than 1, clearly not fulfilling the adiabatic condition.

{Finally, following the path (b) in Fig.\,\ref{fig:harmonicMotion_Phasediagram}, we observe the temporary generation and disappearance of $M_y$ and hence of the $|a_{CP}|$ component in regions with $\mathcal{A} \sim 10$. Interestingly, the magnetization  remains largely periodic under the harmonic fluctuation without inducing strong irreversible transitions. This phenomenon is consistent with the observation in the linear ramping case and can be explained similarly by the first order correction given in Eq.~\eqref{eq:Mx-My-first-order}. In such case, no "true" non-adiabatic transition is induced.}

\subsection{Return to the origin}

An important class of field fluctuations are cyclic where after some time $T$, the field  returns to its original value at $t=0$. Examples include the harmonic field fluctuations considered in the previous section.

As the field returns to its original value, the magnetization generated by the CPMG sequence does in general not recover its original value (neglecting relaxation effects). However, there is an important exception: The CPMG component of the magnetization recovers its initial value if the path of the field fluctuation is confined to an adiabatic region of the phase diagram. This is for instance the case for path $a$ in Fig.\,\ref{fig:harmonicMotion_Phasediagram}. Indeed, the magnetization shown in Fig.\,\ref{fig:harmonicMotion}(a) periodically recovers its full initial value at times when the field returns to its initial value. This is not observed consistently for path $c$, as the path now crosses non-adiabatic regions of the phase diagram.

\begin{figure}
	   \centering
        \adjincludegraphics[width=6.0in, trim={{.15
       \width} {.00\width} {.25\width} {.00\width}},clip]{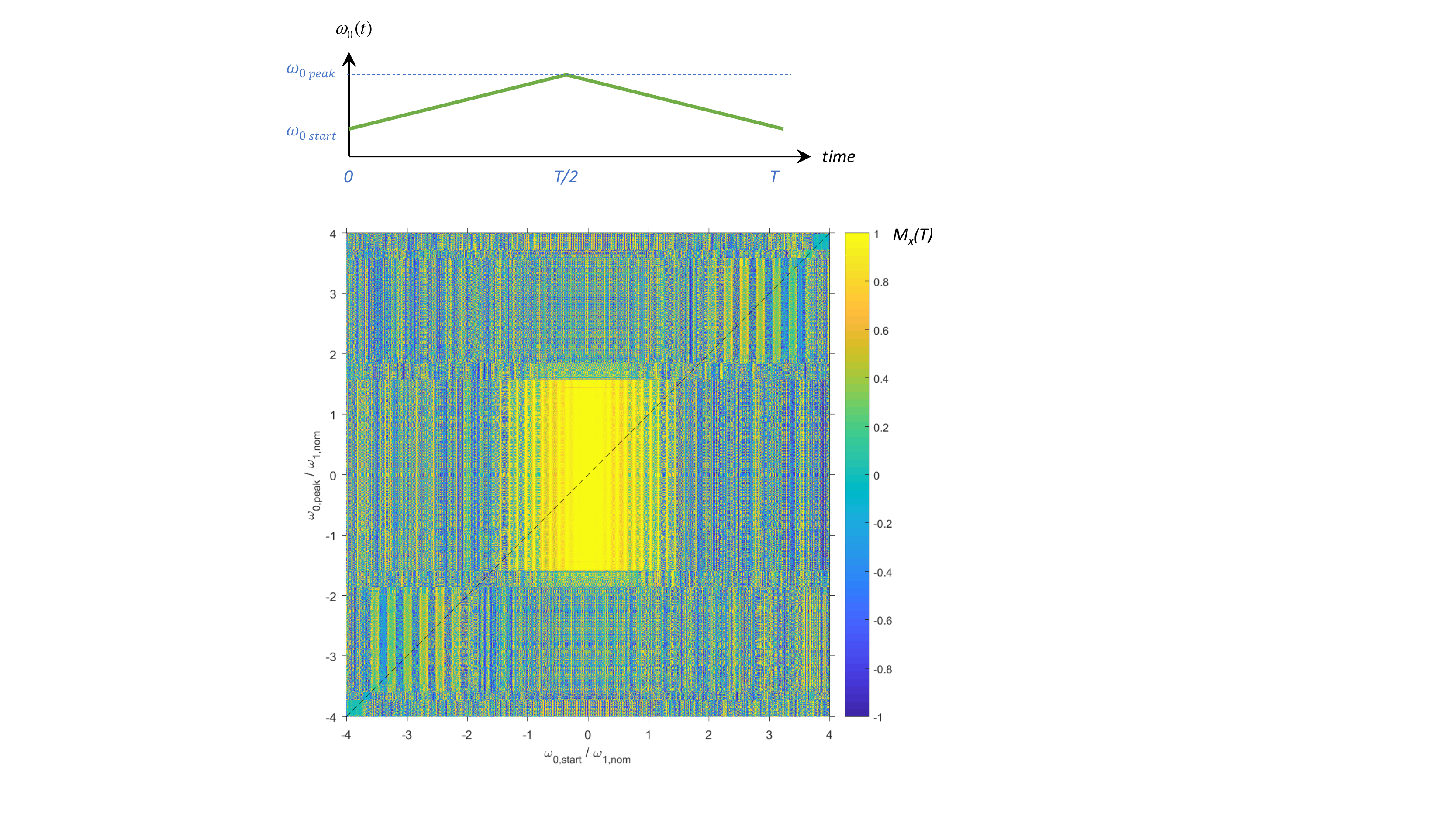}
  	 \caption[Figure 1]{Return to the origin field fluctuations. The top panel shows the $B_0$ field fluctuations considered here: a linear ramp from $\omega_{0,start}$ to $\omega_{0,peak}$ followed by a reverse ramp back to $\omega_{0,start}$. The ramp rate was assumed to be  $\left| \frac{d{{\tilde \omega}_0} } {d\tau} \right| = 10^{-3}$. The main panel shows the resulting in-phase magnetization $M_x(t= T)$ at the end of the ramps as a function of $\omega_{0,start}$ and $\omega_{0,peak}$. The dashed line indicates the diagonal where $\omega_{0,start} = \omega_{0,peak}$ and no field variation occurred during the CPMG sequence. The simulation assumed $t_E / t_{180} = 15$. }
	   \label{fig:Mx_rto}
\end{figure}

\begin{figure}
	   \centering
        \adjincludegraphics[width=6.0in, trim={{.15
       \width} {.05\width} {.10\width} {.05\width}},clip]{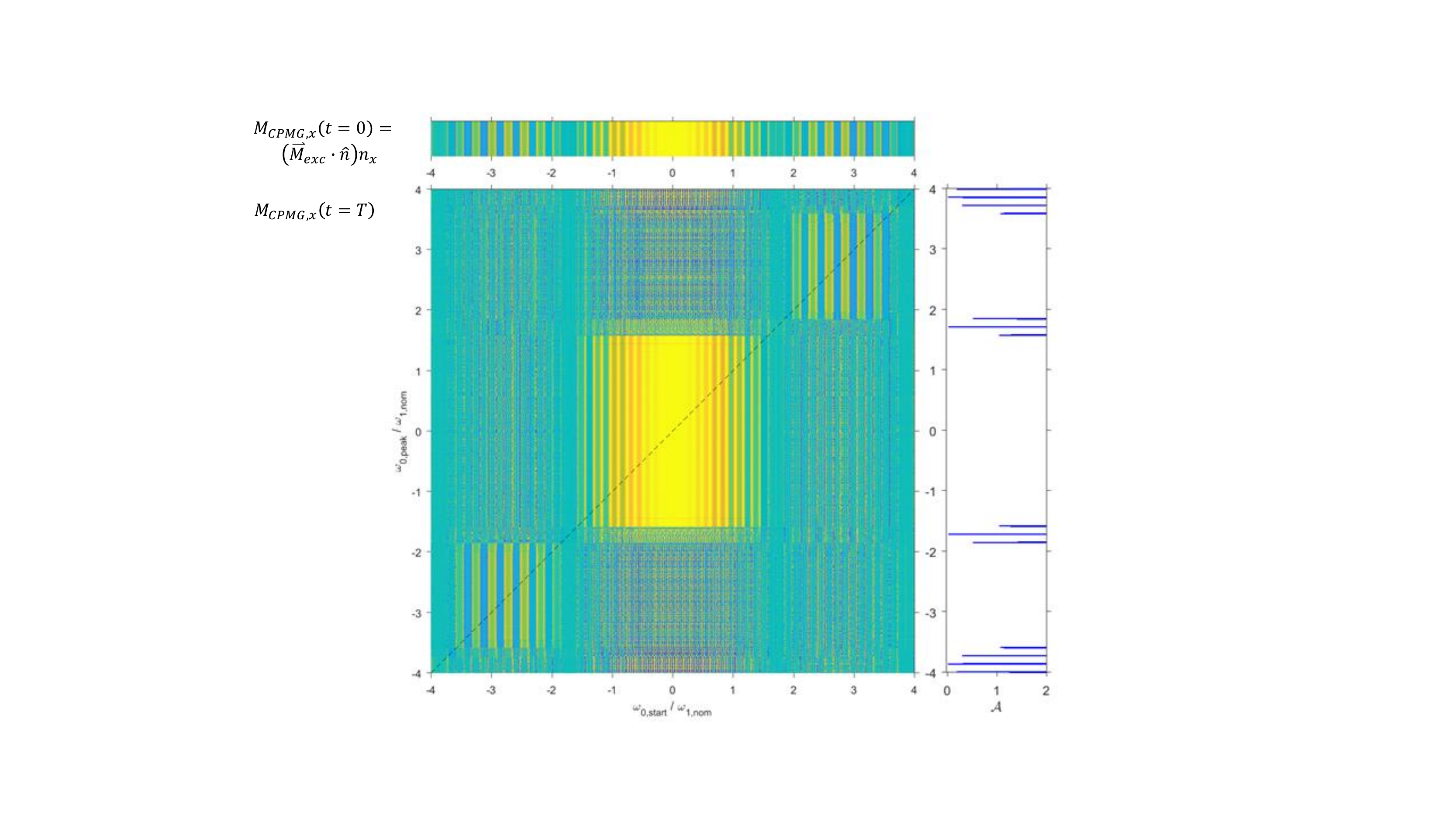}
  	 \caption[Figure 14]{In-phase CPMG component, $M_{CPMG,x}(t= T)$, of the results shown in Fig.\,\ref{fig:Mx_rto}. The right panel shows the adiabaticity parameter $\cal A$ as function of offset frequency and the top panel the initial CPMG magnetization $M_{CPMG,x}(t= 0)$.  }
	   \label{fig:Mx_CPMG_rto}
\end{figure}

We explore this effect more systematically for the case of a bi-linear field variation with results presented in Fig.\,\ref{fig:Mx_rto} and \ref{fig:Mx_CPMG_rto}. Here we assume a linear field ramp from $\omega_{0,start}$ to $\omega_{0,peak}$, followed by an reverse ramp back from  $\omega_{0,peak}$ to $\omega_{0,start}$. We assume that the magnitudes of the ramp rates are fixed at $\left|\frac{d{{\tilde \omega}_0} } {d\tau}\right| = 10^{-3}$. The total duration of the field excursion, $T$, is then $T/t_E = 2 \left|(\omega_{0,peak} - \omega_{0,start})/\omega_{1,nom}\right| / \frac{d{{\tilde \omega}_0} } {d\tau}$.

In Fig.\,\ref{fig:Mx_rto}, we present the transverse magnetization at the end of the ramps, $M_x(T)$, as a function of the starting and peak values of the offset frequency, $\omega_{0,start}$ and $\omega_{0,peak}$, respectively. At time $T$, the offset frequency has returned to its starting value: $\omega_0(T) =  \omega_{0,start}$. The field variation during the ramp is $\omega_{0,peak} - \omega_{0,start}$, i.e. the  vertical distance of the point of interest from the diagonal shown as dashed line.

The results in Fig.\,\ref{fig:Mx_rto} show vertical stripe-like structures in square regions along the diagonal with a superposition of some noise-like features.
The noise-like features are caused by contributions from the CP mode to $M_x(T)$. Even if the fluctuations explore purely adiabatic regions of the phase space, the magnetization of the CP component, $\vec{M}_{CP}(T)$, is highly sensitive to the exact path taken. This is evident from the phase factor $\exp \left\{ -\frac{i}{\hbar} \int_0^t E_k(t') dt' +i \Gamma_k(t) \right\}$ in Eq.\,\eqref{eq:Magn_adia_con}.

For $\omega_{0,start} \neq 0$, CP contributions are generated at the start of the CPMG sequence. Even for fully adiabatic field fluctuations, the total magnetization will then exhibit a noticeable path sensitity through the CP contribution. We can simplify the analysis of Fig.\,\ref{fig:Mx_rto} by analyzing only the CPMG component of $M_x(T)$. In Fig.\,\ref{fig:Mx_CPMG_rto}, we present $M_{CPMG,x}(t= T)$ that was obtained by projecting $\vec{M}(T)$ onto ${\hat n}(T)$.

The square regions along the diagonal now stand out clearly and they are free of noise-like features. Within each square, the magnetization is uniform along the $\omega_{0,peak}$ dimension. The central square extends from $\pm 1.58 \omega_{1,nom}$, whereas the satellite squares are positioned from $\pm\left( 1.85 \,{\rm to}\, 3.59 \right)\omega_{1,nom}$ along both the $\omega_{0,start}$ and $\omega_{0,peak}$ dimensions. These ranges correspond exactly to the adiabatic regions for $\left|\frac{d{{\tilde \omega}_0} } {d\tau}\right| = 10^{-3}$, as shown by the adiabaticity parameter $\cal A$ displayed on the right of Fig.\,\ref{fig:Mx_CPMG_rto}. In these regions, the spin dynamics is fully adiabatic and the CPMG component is fully reversible, i.e. ${\vec M}_{CPMG}(t = T) = {\vec M}_{CPMG}(t = 0)$. Its amplitude is given by $\left( {\vec M}_{exc} \cdot {\hat n} \right) {\hat n}$. The calculations of this initial amplitude of the CPMG component are shown on the top of Fig.\,\ref{fig:Mx_CPMG_rto}.

Outside the square regions, there are still noise-like features present. This indicates that the CPMG magnetization detected at $t=T$ has contributions from the CP mode with at least one mode transition during $T$. In general the paths of the corresponding field fluctuations traverse non-adiabatic regions where the adiabatic condition ${\cal A} \gg 1$ is not fulfilled.

The full simulation of the different field fluctuations shown in Fig.\,\ref{fig:Mx_rto} is computationally expensive. However, the results presented here demonstrates that the key features can be readily obtained from the analytical results with minimal computational effort. The determination of the boundaries of the special square regions requires the calculation of ${\cal A}(\omega_{0,peak})$, a property of a single refocusing cycle. The amplitudes of the magnetization within the special regions in turn requires only the simple calculation of the initial amplitude of the CPMG mode, determined by ${\vec M}_{exc}(\omega_{0,start})$ and ${\hat n}(\omega_{0,start})$.

\section{Conclusion}

We have developed the theoretical framework to analyze the spin dynamics of the CPMG sequence with fluctuating magnetic fields. The analysis is based on the decomposition of the magnetization into eigenmodes of the refocusing cycle. One of the eigenmodes (labeled CPMG mode) is always associated with eigenvalue 1, whereas the other modes (labeled CP modes) have eigenvalues of the form $e^{\pm i \alpha}$, where the angle $\alpha$ depends on the off-resonance condition characterized by the normalized offset frequency ${\tilde \omega}_0$. We introduce the adiabaticity parameter $\cal A$ that is the ratio of the critical velocity $\nu_{0,crit}({\tilde \omega}_0)$ and the normalized rate of change of the magnetic field, $d{\tilde \omega}_0 / d\tau$. The spin dynamics is adiabatic without any transitions between the CPMG and CP modes when the adiabatic condition ${\cal A} \gg 1$ is fulfilled. There are simple analytical solutions in this regime. Otherwise, mode transitions can occur and the response becomes chaotic. Field fluctuations can be characterized by their path in the  ${\tilde \omega}_0 - d{\tilde \omega}_0 / d\tau$ plane to predict the range of the adiabatic regime and the onset of the chaotic behavior associated with the non-adiabatic regime.
The theoretical analysis has been extensively tested with numerical simulations. We have also outlined the extension of this analysis of $B_0$ fluctuations to joint $B_0 - B_1$ fluctuations.

For the standard CPMG sequence, the adiabaticity parameter (and the critical velocities) show pronounced minima at offset frequencies close to $\pm \sqrt{3} \omega_{1,nom}$, $\pm \sqrt{15} \omega_{1,nom}$,...  This behavior is caused by close-by degeneracies between the CPMG and CP modes. It is difficult to pass through the neighborhood of degenerate points without entering the non-adiabatic regime that results in a chaotic response.

The analysis is based on the general properties of the propagator of a single refocusing cycle. Here we considered the simplest possible refocusing cycle containing a single pulse, but this approach is applicable to more general cases that incorporate composite pulses \cite{hurlimann_composite,borneman2010,mandal2013b} or frequency sweeps \cite{garwood2001,schurko2013}. This also opens the possibility to design optimized refocusing pulses that are more robust to $B_0$ and or $B_1$ fluctuations than the standard pulses. The key criterium is to design refocusing sequences that have no singular points close to the range of operations, i.e. to avoid extreme minima in the critical velocities $\nu_{0,crit}$ and $\nu_{1,crit}$.

The present approach can be considered as a generalization of the standard technique of decomposing the magnetization into different coherence pathways \cite{kaiser74,bodenhausen84}. In the standard approach, the magnetization is decomposed into the eigenmodes of free precession. Since every refocusing pulse can induce transitions, it is essential to keep track of the coherence in each refocusing cycle. This makes this approach impractical as the number of relevant coherence pathways grows exponentially with echo number. The current approach
avoids this divergence of coherence pathways. Rather than using the eigenmodes of free precession, we perform here a decomposition into the eigenmodes of the refocusing cycle. We only have to consider transitions between the different adiabatic regions that are separated by regions where the adiabatic condition ${\cal A} \gg  1$ is not fulfilled. The adiabatic regions typically extend over a large number of refocusing cycles, which results in a large decrease of relevant coherence pathways.

The transition rates between the different adiabatic regimes are highly sensitive to the external parameters such as ramp rates. This implies that when transitions between the CP and CPMG levels occur, the outcome of an experiment on a single spin is difficult to predict in a robust manner. There is no problem calculating the transition rates numerically, but it will be challenging to determine the actual external parameters sufficiently accurately.

\appendix

\section{Transition between CPMG and CP modes in the continuous limit}
\label{app:B}

In this Appendix, we will provide a brief review on the derivation of Eq.~\eqref{eq:mod_B_eff}, following closely those in Ref.~\cite{Berry2009}, and then discuss how to use the inverted form of Eq.~\eqref{eq:mod_B_eff} to compute the transition rate between different eigenstates.

\subsection{Derivation of Eq.~\eqref{eq:mod_B_eff}}

Let us consider an arbitrary time-dependent Hamiltonian, $\mathcal{H}_0(t)$, that has the instantaneous eigenstates, $|u_{\ell}(t) \rangle$, with energies $E_{\ell}(t)$. In the adiabatic approximation, the evolution of states is given by
\begin{equation}
|\psi_{\ell}(t)\rangle = \exp \left\{ -\frac{i}{\hbar}\int dt'E_\ell(t') - \int dt' \langle u_{\ell}(t')| \partial_{t'} u_{\ell}(t')\rangle  \right\} |u_{\ell}(t) \rangle.
\label{eq:adia_evo_arb_H}
\end{equation}
The first integral inside the exponential is simply the dynamic phase while the second one related to the geometric phase accumulated by the time dependent state. This adiabatic time evolution of all states can be defined by a time-dependent unitary operator expressed by
\begin{equation}
U(t)= \sum_{\ell} \exp \left\{ -\frac{i}{\hbar}\int dt'E_\ell(t') - \int dt' \langle u_{\ell}(t')| \partial_{t'} u_{\ell}(t')\rangle  \right\} |u_{\ell}(t) \rangle \langle u_{\ell}(0)|.
\end{equation}
Interestingly, there exists a modified Hamiltonian $\mathcal{H}_{mod}(t)$ that gives rise to this exact time-dependent unitary operator in the general case, not just in the adiabatic limit. It can be shown that this modified Hamiltonian $\mathcal{H}_{mod}(t)$ can be formally expressed as:
\begin{equation}
 \mathcal{H}_{mod}= i \hbar (\partial_t U(t))U^{\dagger}(t),
 \label{eq:H_formal}
\end{equation}
where the symbol $\dagger$ indicates the Hermitian conjugate of the operator. Explicitly, this constructed Hamiltonian can be written as
\begin{equation}
\begin{split}
\mathcal{H}_{mod}(t)=& \sum_{\ell}  E_\ell(t)  |u_{\ell} \rangle\langle u_{\ell}| +i \hbar \sum_{\ell} \left( |\partial_t u_{\ell} \rangle \langle u_{\ell}| - \langle u_{\ell}| \partial_t u_{\ell}\rangle | u_{\ell} \rangle \langle u_{\ell}| \right)
\\
\equiv& \mathcal{H}_{0}(t) +\mathcal{H}_1(t).
\end{split}
\label{eq:H_formal_1}
\end{equation}
Here, all the bra and ket states are evaluated at time $t$. The first summation simply gives $\mathcal{H}_{0}(t)$ as the original Hamiltonian while the second summation is grouped into $\mathcal{H}_1(t)$ as the additional term required for driving eigenstates to follow the adiabatic evolution in Eq.~\eqref{eq:adia_evo_arb_H}.

The additional part of Hamiltonian $\mathcal{H}_1(t)$ can be rewritten into the form
\begin{equation}
\label{eq:H_1}
\mathcal{H}_1(t)= i\hbar \sum_{\ell'\neq \ell} \sum \frac{|u_{\ell'}\rangle \langle u_{\ell'}| \partial_t H_0 |u_{\ell}\rangle \langle u_{\ell}|}{E_{\ell}-E_{\ell'}},
\end{equation}
with the aid of the identity
\begin{equation}
\langle u_{\ell'} | \partial_t u_{\ell}\rangle =\frac{ \langle u_{\ell'}| \partial_t H_0 |u_{\ell}\rangle}{E_{\ell}-E_{\ell'}}.
\end{equation}
It is worthwhile to point out that the above procedure can be utilized to construct the exactly driving Hamiltonian $\mathcal{H}_{mod}(t)$ for any time-dependent Hamiltonian $\mathcal{H}_0(t)$ without generating transition between them.

We can now compute the exact driving Hamiltonian $\mathcal{H}_{mod}(t)$ for the original time-dependent Hamiltonian given in Eq.~\eqref{eq:H_con}, $\mathcal{H}_{0}(t) = \mathcal{H}_{ave}(t)$. From the expression of $H_1$ in Eq.~\eqref{eq:H_1}, we have
\begin{equation}
\label{eq:H_1-1}
\mathcal{H}_1(t)= i \hbar^2 \gamma \left(\partial_{t} \mathcal{\boldsymbol{B}}_{\mathit ave}(t)\right) \cdot \sum_{\ell'\neq \ell} \sum \frac{|u_{\ell'}\rangle \langle u_{\ell'}| \vec{\mathcal{S}} |u_{\ell}\rangle \langle u_{\ell}|}{E_{\ell}-E_{\ell'}}
\end{equation}
After some algebra following Ref.~\cite{Berry2009}, one can cast Eq.~\eqref{eq:H_1-1} into the concise expression:
\begin{equation}
\mathcal{H}_1(t)= \frac{\hbar}{|\mathcal{\boldsymbol{B}}_{\mathit ave}(t)|^2} \mathcal{\boldsymbol{B}}_{\mathit ave}(t)\times \partial_{t} \mathcal{\boldsymbol{B}}_{\mathit ave}(t)=\hbar \hat{n}(t)\times \partial_t \hat{n}(t) .
\end{equation}
The exact driving Hamiltonian can now be written as:
\begin{equation}
\mathcal{H}_{\mathit mod}(t)=\mathcal{H}_0(t)+\mathcal{H}_1(t)= \hbar \left[ \gamma \mathcal{\boldsymbol{B}}_{\mathit ave} + \hat{n}(t)\times \partial_t \hat{n}(t) \right] \cdot  \vec{\mathcal{S}} \equiv \hbar \gamma \mathcal{\boldsymbol{B}}_{\mathit mod}\cdot  \vec{\mathcal{S}}.
\label{eq:H_exact_drive_fianl}
\end{equation}
This gives the modified effective B-field given in Eq.~\eqref{eq:mod_B_eff}.

\subsection{Inverted interpretation and transition rate}

The exact driving Hamiltonian $\mathcal{H}_{\mathit mod}(t)$ in Eq.~\eqref{eq:H_exact_drive_fianl} is derived knowing both the original Hamiltonian $\mathcal{H}_0(t)$ and its corresponding adiabatic evolution states. Interestingly, for a spin coupled to the magnetic field, it also provides a way to compute the exact evolution driven by the original Hamiltonian, $\mathcal{H}_{\mathit ave}(t)$ beyond the adiabatic limit, i.e. compute the exact time dependent eigenstates driven by the original Hamiltonian, $\mathcal{H}_{\mathit ave}(t)$.
More precisely,  Eq.~\eqref{eq:mod_B_eff} can be used to read
\begin{equation}
\gamma \mathcal{\boldsymbol{B}}_{\mathit ave}(t) = \left[\gamma \mathcal{\boldsymbol{B}}_{\mathit eff}(t) + \hat{n}_{\mathit eff}(t)\times \partial_t \hat{n}_{\mathit eff}(t) \right],
\label{eq:inverted_B_eff}
\end{equation}
where $\hat{n}_{\mathit eff}(t)\equiv\mathcal{\boldsymbol{B}}_{\mathit eff}(t)/\left|\mathcal{\boldsymbol{B}}_{\mathit eff}(t)\right|$. We are searching for an effective field $\mathcal{\boldsymbol{B}}_{\mathit eff}(t)$ such that its modified field $\mathcal{\boldsymbol{B}}_{\mathit mod}(t)$ is identical to the original field $\mathcal{\boldsymbol{B}}_{\mathit ave}(t)$. The direction vector $\hat{n}_{\mathit eff}(t)$ then
corresponds to the exact eigenstate driven by $\mathcal{H}_{\mathit ave}(t)$. The task now reduces to solving the coupled differential equations Eq.~\eqref{eq:inverted_B_eff} for each component of $\mathcal{\boldsymbol{B}}_{\mathit eff}$ and $\hat{n}_{\mathit eff}(t)$. The general form for the magnetization has then the form of Eq.~\eqref{eq:Magn_adia_con}, even outside the adiabatic limit. In particular for the important case when ${\vec M}_{exc} = \hat{n}_{\mathit eff}(t=0)$, the solution reduces to the simple form ${\vec M}(t) = \hat{n}_{\mathit eff}(t)$.

In principle, Eq.~\eqref{eq:inverted_B_eff} applies to any spin system coupled to a well-defined and continuous magnetic field. For the specific problem on our hand, there are two caveats worth to keep in mind. First, the effective magnetic field, $\mathcal{\boldsymbol{B}}_{\mathit ave}(t) = (\alpha(t)/t_E) \hat{n}(t)$, is derived from the analysis of the rotation operation on the magnetization at each refocusing cycle. However, the direction of the rotation axis is only defined modulo a factor of $-1$. Following the convention in the main text, we constrained the effective rotation angle to $0\leq \alpha\leq \pi$. This can lead to discontinuities in the direction of $\mathcal{\boldsymbol{B}}_{\mathrm ave}$  when $\alpha$ approaches $\pi$ and the corresponding rotating direction $\hat{n}(t)$ switches between opposite directions.
Hence, for using Eq.~\eqref{eq:inverted_B_eff} to evaluate the evolution of state or transition rate with time-dependent fields, we need to solve it piece-wisely between the discontinuous points and to project eigenstates into the new basis at the discontinuous points.

Second, we treat the effective magnetic field, $\mathcal{\boldsymbol{B}}_{\mathit ave}(t)$ as the continuous representation of the discrete rotations generated by each refocusing cycle. When the applied magnetic field $B_0$ and/or the pulsed field $B_1$ vary too fast, this continuous approximation may not be  fully justified. Although the qualitative properties, such as occurrences of mode transitions, can still be understood, one should not expect to obtain quantitative results from the continuous formalism in those cases.

\begin{figure}
	\centering
	\adjincludegraphics[width=3.75 in, trim={{.00
			\width} {.0\width} {.00\width} 0},clip]{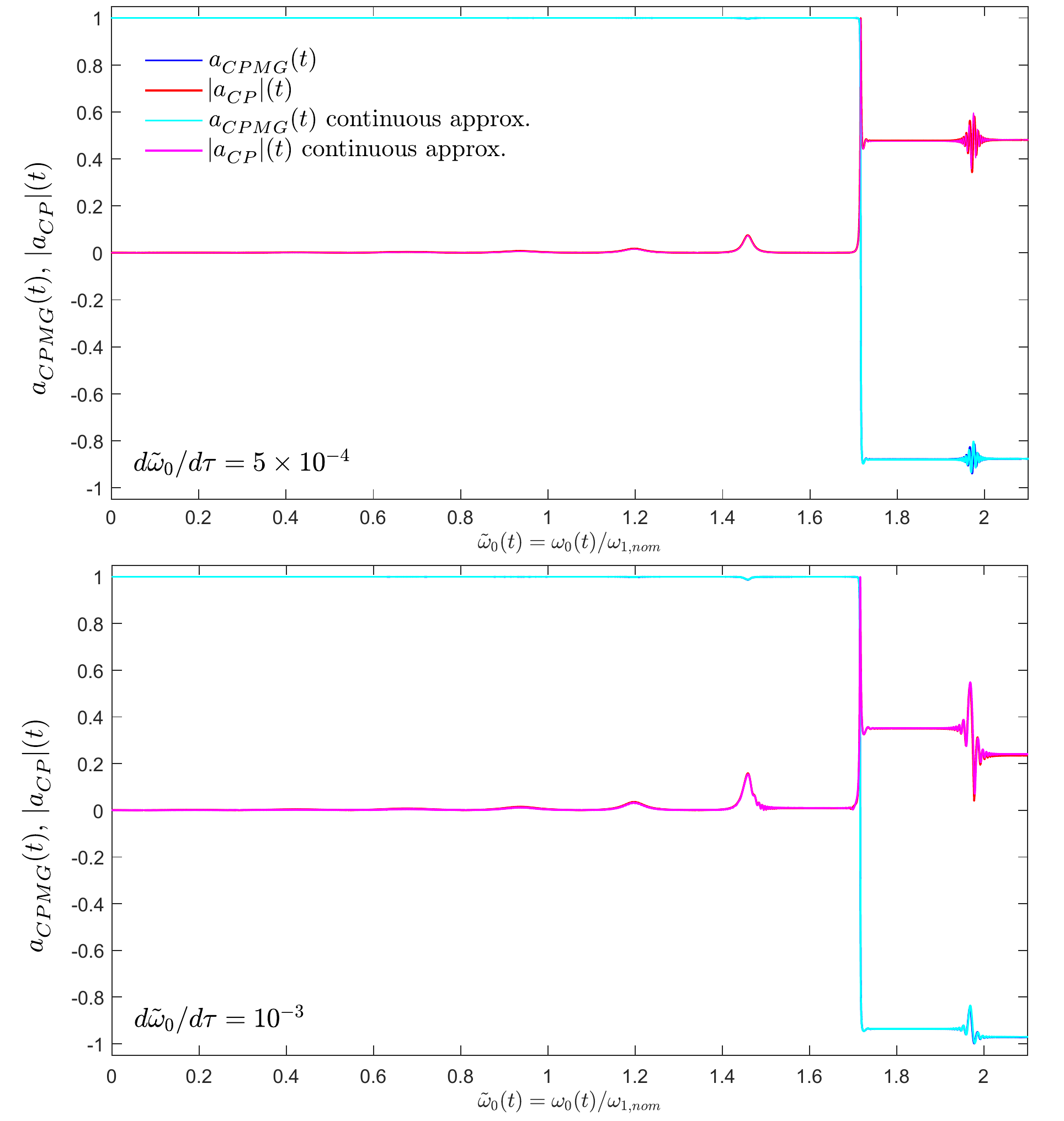}
	\caption{The CPMG amplitudes and CP magnitudes obtained from the direct spin dynamic simulation (red and blue curves) and from continuous approximation (cyan and magenta curves). Two offset frequency increasing rate, $d\tilde{\omega}_0/d\tau = 5\times 10^{-4}$ and $10^{-3}$, and $t_E / t_{180} = 8$ are used for simulation.}
	\label{fig:con_dis_com}
\end{figure}

In Sec.~\ref{sec:Results}, we have used Eq.~\eqref{eq:inverted_B_eff} together with the first order perturbation calculation to obtain Eq.~\eqref{eq:Mx-My-first-order} that explains the variations of the magnetization $|M_y|$ in the linear ramping process up to the non-adiabatic region. Eq.~\eqref{eq:inverted_B_eff}, however, is quite general and allows us to faithfully capture the transition between CPMG and CP modes
in the non-adiabatic region. We have tested this approach of calculating the spin dynamics based on solving Eq.~\eqref{eq:inverted_B_eff} by comparing the results with those based on the standard numerical integration of the Bloch equation used in generating the results presented in section \ref{sec:Results}. Figure~\ref{fig:con_dis_com} shows comparisons for simulations of linear field fluctuations with two different ramp rates. The calculations assumed that at the start of the ramp $\tilde{\omega}_0(t=0)=0$ and that the magnetization was initially all in the CPMG mode, $a_{CPMG}(t=0)=1$. The refocusing cycle duration is eight times of the pulse width, $t_{E}/t_{180}=8$. Figure~\ref{fig:con_dis_com} displays the evolution of the amplitudes of CPMG and CP modes calculated by the two approaches. The results from the two approaches nearly coincide, thus validating the alternative method based on the continuous approximation and Eq.~\eqref{eq:inverted_B_eff}. Only small discrepancies can be observed. All the primary features and even the quantitative values of amplitudes of CPMG and CP modes from the continuous approximation are comparable to those obtained from direct spin dynamics simulation.

\newpage

\section*{References}
\bibliographystyle{elsarticle-num}
\bibliography{NMR_MDH}

\end{document}